\documentclass[12pt]{article}
\usepackage{citesort,bibmods}
\usepackage{amssymb,latexsym}
\usepackage[final]{graphicx}
\usepackage{float}
\usepackage{pstricks}
\usepackage{a4}

\def\greaterthansquiggle{\raise.3ex\hbox{$>$\kern-.75em\lower1ex\hbox{$\sim$}}}
\def\lessthansquiggle{\raise.3ex\hbox{$<$\kern-.75em\lower1ex\hbox{$\sim$}}}

\newcommand{\UM}{UM}

\newcommand{\beq}{\begin{equation}}

\newcommand{\eeq}{\end{equation}}
\newcommand{\beqa}{\begin{eqnarray}}
\newcommand{\eeqa}{\end{eqnarray}}
\newcommand{\beqan}{\begin{eqnarray*}}
\newcommand{\eeqan}{\end{eqnarray*}}
\newcommand{\ba}{\begin{array}}
\newcommand{\ea}{\end{array}}

\newcommand{\ol}{\overline}

\def\nz{\ifmmode {I\hskip -3pt N} \else {\hbox {$I\hskip -3pt N$}}\fi}
\def\zz{\ifmmode {Z\hskip -4.8pt Z} \else
       {\hbox {$Z\hskip -4.8pt Z$}}\fi}
\def\qz{\ifmmode {Q\hskip -5.0pt\vrule height6.0pt depth 0pt
       \hskip 6pt} \else {\hbox
       {$Q\hskip -5.0pt\vrule height6.0pt depth 0pt\hskip 6pt$}}\fi}
\def\rz{\ifmmode {I\hskip -3pt R} \else {\hbox {$I\hskip -3pt R$}}\fi}
\def\cz{\ifmmode {C\hskip -4.8pt\vrule height5.8pt\hskip 6.3pt} \else
       {\hbox {$C\hskip -4.8pt\vrule height5.8pt\hskip 6.3pt$}}\fi}
\def\au{{\setbox0=\hbox{\lower1.36775ex\hbox{''}\kern-.05em}\dp0=.36775ex\hs
kip0pt\box0}}
\def\ao{{}\kern-.10em\hbox{``}}

\newtheorem{Theorem} {Theorem} [section]
\newtheorem{Corollary} [Theorem] {Corollary}
\newtheorem{Lemma} [Theorem] {Lemma}
\newtheorem{Proposition} [Theorem] {Proposition}

\newtheorem{example}[Theorem]{Example}

{\catcode `\@=11 \global\let\AddToReset=\@addtoreset}
\AddToReset{equation}{section}

\newcommand{\novec}[1]{{#1}}

\def\scri{\hbox{${\cal J}$\kern -.645em {\raise
      .57ex\hbox{$\scriptscriptstyle (\ $}}}}

\newcommand{\eq}[1]{(\ref{#1})}

\newcommand{\commentout}[1]{}

\newcommand{\ee}{\end{equation}}
\newcommand{\bea}{\begin{eqnarray}}
\newcommand{\eea}{\end{eqnarray}}
\newcommand{\beaa}{\begin{eqnarray*}}
\newcommand{\eeaa}{\end{eqnarray*}}

\newcommand{\R}{{\mathbb R}}
\newcommand{\Na}{{\cal{ N}^+}}

\begin{document}

\title{``Nowhere'' differentiable horizons \\
}
\author{Piotr T.\ Chru\'sciel\thanks{%
On leave of absence from the Institute of Mathematics, Polish Academy of
Sciences, Warsaw. Supported in part by KBN grant \# 2P302 095 06. \emph{%
E--mail}: Chrusciel@Univ-Tours.fr} \\
D\'epartement de Math\'ematiques\\
Facult\'e des Sciences\\
Parc de Grandmont\\
F37200 Tours, France\\
\\
Gregory J. Galloway\thanks{Supported in part by NSF grant \# DMS-9204372.
\emph{E--mail}:
galloway@math.miami.edu}\\
Department of Mathematics and Computer Science\\
University of Miami\\
Coral Gables FL 33124, USA}
\maketitle

\begin{abstract}
  It is folklore knowledge amongst general relativists that
  horizons are
well behaved, continuously differentiable hypersurfaces except
  perhaps on a negligible subset one needs not to bother with. We show
  that this is not the case, by constructing a Cauchy horizon, as well
  as a black hole event horizon, which contain no open subset on which
  they are differentiable.
\end{abstract}

\section{Introduction}

\label{introduction}

In various works where Cauchy horizons or event horizons occur it is assumed
that those objects have a fair amount of differentiability, except perhaps
for a well behaved lower dimensional subset thereof. A standard example is
the proof of the area theorem for black holes \cite[Proposition 9.2.7]{HE},
where something close to $C^2$ differentiability ``almost everywhere'' of
the event horizon seems to have been assumed. An extreme example is the
rigidity theorem for black-holes \cite{Ha1,HE}, where it is assumed that the
event horizon is an analytic submanifold of space--time.
Other examples
include discussions of the structure of Cauchy horizons (e.g., \cite[p.
295--298]
{HE}, \cite{Hawking:timemachines,Tipler,Galloway:global,%
Newman:noreturn,Beem:Cauchy}). 
It seems reasonable to raise the
question, whether there are any reasons to expect that the horizons
considered in the above quoted papers should have the degree of
differentiability assumed there. While we do not have an answer to that
question, in this paper we construct examples of horizons the degree of
differentiability of which is certainly not compatible with the
considerations in the papers previously quoted.

Our examples are admittedly very artificial and display various
pathological features (\emph{e.g.}, an incomplete null infinity). One
would hope that the behavior described here can not occur, in the case
of event horizons in ``reasonable'' asymptotically flat space--times
(\emph{e.g.}, in vacuum and maximal globally hyperbolic space--times),
or for compact Cauchy horizons in spatially compact space--times. While
this expectation might well turn out to be true, the examples here
show that we have no \emph{a priori} reasons to expect that to be
correct, and some new insights are necessary in those cases to
substantiate such hopes. It is, moreover,
clear from our results below that the kind of behaviour we obtain is a
\emph{generic} property of horizons as considered here, in an
appropriate $C^{0,1}$ topology on the collection of sets. We shall,
however, not attempt to formalize this statement. We note that
a criterion for stability of differentiability, for a rather special
class of compact Cauchy horizons, has been given in
\cite{ChIs:stability}.

Let us recall some facts about, say, Cauchy horizons in globally hyperbolic
space--times $(M,g)$. Let $\Omega$ be an open subset of a Cauchy surface $%
\Sigma\subset M$. Then \cite[Prop. 6.3.1]{HE} \cite[Lemma 3.17]
{PenroseDiffTopo}, ${\mathcal{H}}^+(\Omega)$ is a Lipschitz topological
submanifold of $M$.
It then follows from a theorem of
Rademacher \cite[Theorem 3.1.6]{FedererMeasureTheory} that ${\mathcal{H}}%
^+(\Omega)$ is differentiable almost everywhere. (Recall that in local
coordinates $x^\mu$ in which $x^0$ is a time function ${\mathcal{H}}%
^+(\Omega)$ can be written as a graph $x^0=f(x^i)$, and then the ``almost
everywhere'' above refers to the Lebesgue measure $\mathrm{d}%
x^1\wedge\ldots\wedge \mathrm{d} x^n$.) This shows that there is some
smoothing associated with Cauchy horizons: consider a set $K$ such that $%
\partial K$ is a nowhere differentiable curve. (Take, e.g., a Mandelbrot set
in ${\mathbb R}^2$, considered as the Cauchy surface for the
three-dimensional Minkowski space--time ${\mathbb R}^{2,1}$.)
Even in such a
``bad'' situation, for almost all (in the sense of Lebesgue measure on
$\R$)
later-times $t$ the boundary of $D^+(K)$
intersected with the plane $t= \mbox{const}$ will be a set the boundary of
which will be a curve which will be differentiable almost everywhere.
It could be the case that some more smoothing occurs, so that perhaps ${%
\mathcal{H}}^+(K)\cap \{t=\mbox{const}\}$ will have nicer differentiability
properties than what follows from the above considerations. The example we
construct here shows that this does not happen. More precisely, we show the
following:

\begin{Theorem}
\label{T1} There exists a connected set $K\subset {\mathbb R}%
^2=\{t=0\}\subset {\mathbb R}^{2,1}$, where ${\mathbb R}^{2,1}$ is the $2+1$
dimensional Minkowski space--time, with the following properties:

\begin{enumerate}
\item  The boundary $\partial K=\bar K\setminus \mathrm{int}\,K$ of
  $K$ is a
connected, compact, Lipschitz topological submanifold of ${\mathbb R}^2$. $K$
is the complement of a compact set of ${\mathbb R}^2$.

\item  There exists no open set $\Omega \subset {\mathbb R}^{2,1}$ such that
$\Omega \cap {\mathcal{H}}^{+}(K)\cap \{0<t<1\}$ is a differentiable
submanifold of ${\mathbb R}^{2,1}$.
\end{enumerate}
\end{Theorem}

\begin{figure}[htbp]
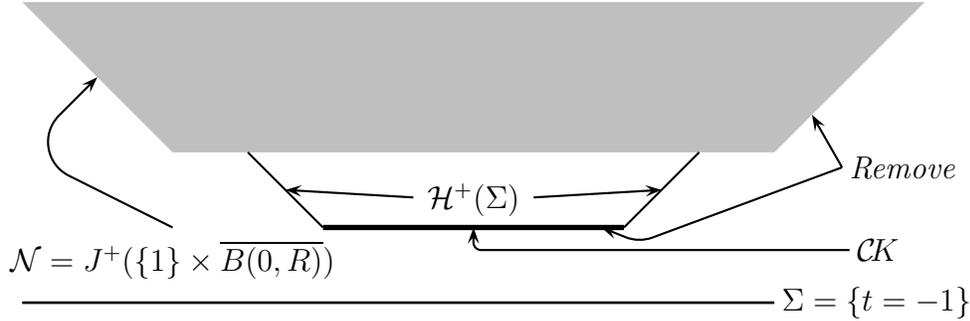

\pspicture(-6,-1)(6,3)
\centering
\rput[b](0,0.15){${\cal H}^+(\Sigma)$}
\psline{<-}(-2.5,0.5)(-0.8,0.4)
\psline{<-}(2.5,0.5)(0.8,0.4)
\psline[linewidth=1pt](-6,-1)(4,-1)
\psline[linewidth=2pt](-2,0)(2,0)
\psline*[linecolor=lightgray](-6,3)(-4,1)(4,1)(6,3)
\uput[270](-4,0){${\cal N}=J^+ (\{1\}\times \ol{B(0,R)})$}
\psline[linearc=.5]{<-}(-5,2)(-6,1)(-4,0)
\psline(-3,1)(-2,0)
\psline(3,1)(2,0)
\rput[l](5,0.8){\em Remove}
\psline[linearc=.5]{<-}(1.7,0)
(2.2,-0.2)(4.9,0.8)
\psline{<-}(4.5,1.5)(4.9,0.8)
\rput[l](4.1,-1){$\Sigma=\{t=-1\}$}
\rput[l](5.1,-0.3){${\cal C}\!K$}
\psline[linearc=.1]{<-}(0,0)(0,-0.3)(5,-0.3)
\endpspicture

\caption{The space--time $\hat M$.}
\label{F:1}
\end{figure}

Let us show that Theorem \ref{T1} implies the existence of ``nowhere''
differentiable (in the sense of point 2 of Theorem \ref{T1}) Cauchy horizons $%
{\mathcal{H}}^+(\Sigma)$ in stably causal asymptotically flat space--times,
with well behaved space--like surfaces $\Sigma$. (Our example satisfies even
the vacuum Einstein equations, as the space--time is flat.) To do that,
consider the complement $\mathcal{C}\! K$ of $K$ in ${\mathbb R}^2$, and
consider the set $J^+(
{\mathcal{C}\! K})$.
As $
{\mathcal{C}\! K}$ is compact, it follows from global hyperbolicity of
${\mathbb R}%
^{2,1} $ that $K^{\prime}\equiv J^+(
{\mathcal{C}\! K})\cap\{t=1\}$
is compact, hence there exists $R$ such that $K^{\prime}$ is a subset of the
open ball $B(0,R)$ of radius $R$ centered at the origin of ${\mathbb R}^2$.
Let $\mathcal{N}=J^+ (\{1\}\times \overline{B(0,R)})$ and set $\hat M={%
\mathbb R}^{2,1}\setminus (\mathcal{N}\cup \mathcal{C}\! K)$, equipped with
the obvious flat metric coming from ${\mathbb R}^{2,1}$ (\emph{cf.}\ Figure
\ref{F:1}).
Consider the space--like surface $\Sigma=\{t=-1\}\subset \hat M$. It is easily
seen that the Cauchy horizon ${\mathcal{H}}^+(\Sigma;\hat M)$ of $\Sigma$ in
$\hat M$ coincides with the intersection of the Cauchy horizon ${\mathcal{H}}%
^+(K)$ in ${\mathbb R}^{2,1}$ with the ``slab'' $\{0<t<1\}\subset {\mathbb R}%
^{2,1}$. Hence there exists no open subset of ${\mathcal{H}}^+(\Sigma;\hat
M) $ such that this subset is a differentiable submanifold of $\hat M$. We
note that the construction guarantees global hyperbolicity of $M$.

\begin{figure}[htbp]
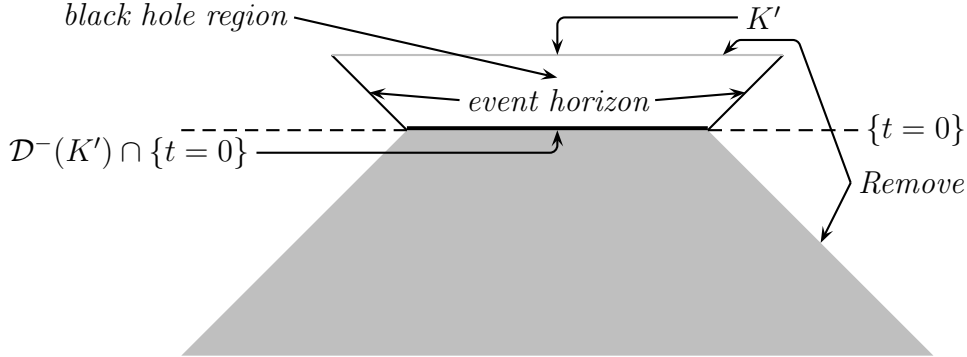

\centering
\pspicture(-6.5,-4)(5.5,3)
\centering
\rput[b](0,0.25){\em event horizon}
\psline{<-}(-2.5,0.5)(-1.3,0.4)
\psline{<-}(2.5,0.5)(1.3,0.4)
\psline(-3,1)(-2,0)
\psline(3,1)(2,0)
\psline[linecolor=lightgray](-3,1)(3,1)
\psline[linewidth=3pt](-2,0)(2,0)
\psline*[linecolor=lightgray](-5,-3)(-2,0)(2,0)(5,-3)
\rput[l](4,-0.7){\em Remove}
\psline{<-}(3.5,-1.5)(3.9,-0.7)
\psline[linearc=.1]{<-}(2.2,1)(2.4,1.2)(3.2,1.2)(3.9,-0.7)
\rput[r](-3.6,1.5){\em black hole region}
\psline{->}(-3.5,1.5)(0,0.7)
\rput[l](2.5,1.5){$K'$}
\psline[linearc=.1]{<-}(0,1)(0,1.5)(2.4,1.5)
\rput[r](-4.1,-0.3){${\cal D}^-(K')\cap \{t=0\}$}
\psline[linearc=.1]{<-}(0,0)(0,-0.3)(-4,-0.3)
\psline[linestyle=dashed](-5,0)(-2,0)
\psline[linestyle=dashed](2,0)(4,0)
\rput[l](4.1,0){$\{t=0\}$}
\endpspicture
\caption{A flat space--time with a black hole region.}
\label{F:2}
\end{figure}

Let us show that Theorem
\ref{T1} also implies the existence of black hole space--times with
``nowhere'' differentiable
event horizons (``nowhere'' in the sense of point 2 of Theorem \ref{T1}).
The construction
should be clear from Figure~\ref{F:2},
and can be formally described as follows:
Construct first a compact set $K'\subset \R^3=\{t=1\}\subset \R^{2,1}$
with the property that the past Cauchy horizon $\partial {\cal
D}^-(K')$ of $K'$ is ``nowhere differentiable'' for $0\le t\le 1$. This
is easily achieved using the same construction as below, by adding
{\em exterior} ripples to the disc $B(0,2)$ of radius $2$ centered at $0$.
Consider the space--time $\tilde M$
obtained by removing from the three--dimensional Minkowski space--time the
set $K'$
together with
$J^-(\overline{{\cal D}^-(K')\cap \{t=0\}})$.
%
Clearly those points in $\tilde M$ which are in $\overline{{\cal D}^-(K')}$
cannot be seen from the usual future null infinity $%
\hbox{${\cal J}$\kern -.645em {\raise
      .57ex\hbox{$\scriptscriptstyle (\ $}}}^+({\tilde M})$ of $\tilde M$
(which coincides with that of ${\mathbb R}^{2,1}$), so that
$\overline{{\cal D}^-(K';\tilde M)}$
lies in the black hole region of $\tilde M$. It
is also easily seen that the black hole horizon $\mathcal{B}$ in $\tilde M$
coincides with ${\mathcal{H}}^-(K')\cap\{0<t<1\}$. It follows again that
there exists no open subset of $\mathcal{B}$ which is a differentiable
submanifold of $\tilde M$. We note that $\tilde M$ has a complete $%
\hbox{${\cal J}$\kern -.645em {\raise
      .57ex\hbox{$\scriptscriptstyle (\ $}}}^+$ and is stably causal, but is
not globally hyperbolic, moreover the domain of outer communications is not
globally hyperbolic either. We could obtain a globally hyperbolic
space--time \emph{and} a globally hyperbolic domain of outer communications
by further removing the set $J^+ (
K')\cap\{t>1\}$
from $\tilde M$, but this would lead to an incomplete $%
\hbox{${\cal J}$\kern -.645em {\raise
      .57ex\hbox{$\scriptscriptstyle (\ $}}}^+$.%

This paper is organized as follows. In Section \ref{SP} we present the
construction of $K$.  While all the assertions made there can be seen,
with some effort, by inspection of the pictures, it seems difficult to
give a rigorous proof based on that. For that reason we develop, in
the following sections, a framework which allows us to handle various
problems one needs to address in the construction presented in Section
\ref{SP}. We believe that the results obtained in those sections are
of independent interest, and shed light on various properties of
Cauchy horizons. In particular the notion of the distribution of
semi--tangents introduced below seems to be rather useful in
understanding the structure of horizons. In Section \ref{S3} we
analyze
continuity properties of the map which to a set $K\subset\Sigma$, where $%
\Sigma$ is a Cauchy surface in a globally hyperbolic space--time, assigns its
Cauchy horizon ${\mathcal{H}}^+(K)$, in various topologies. Some related
results can be found in \cite{Beem:Cauchy}. In Section \ref{S4}
we introduce the notion of the distribution of semi-tangents to a Cauchy
horizon, which plays a crucial role in our analysis. We also study some
properties of that distribution. In Section \ref{S5} we analyze continuity
properties of the map which to a set $K\subset\Sigma$ assigns its
distribution of semi-tangents ${\mathcal{N}}{}^+(K)$. We show that this map
is continuous both in a ``clustering'' topology and in a Hausdorff-distance
topology. In the concluding section we briefly discuss how our
construction generalizes to higher dimensional Minkowski space--times,
or, in fact, to any globally hyperbolic space--times.

\section{Convergence Properties of Cauchy Horizons}

\label{S3}

In this section, and subsequent sections, we establish some convergence
properties of Cauchy horizons. For this purpose we make use of the notion of
Hausdorff convergence.

Let $X$ be a metric space with distance function $d$. For any subset $%
A\subset X$, let $U_{\varepsilon}(A)$ be the $\varepsilon$-neighbourhood
of $A$,
\[
U_{\varepsilon}(A)=\{x\in X:\; d(x,A)<\varepsilon\}
\]
where
\[
d(x,A)=\inf_{y\in A} d(x,y).
\]
For subsets $A$, $B\in X$, the Hausdorff distance between $A$ and $B$,
denoted $D(A,B)$, is defined as follows,
\[
D(A,B)=\inf\{\varepsilon>0:\; A\subset U_{\varepsilon}(B)\mbox{ and }%
B\subset U_{\varepsilon}(A)\}.
\]
Then, a sequence of subsets $A_n\subset X$ is said to (Hausdorff) converge
to a subset $A\subset X$ provided $D(A_n,A)\rightarrow 0$ as $n\rightarrow
\infty$. Note that $D(A_n,A)\rightarrow 0$ if and only if $\displaystyle %
\sup_{x\in A_n} d(x,A)\rightarrow 0$ and $\displaystyle \sup_{y\in A}
d(y,A_n)\rightarrow 0$.

There is an auxilliary convergence notion that turns out to be useful in our
study, as well. Let $\{A_n\}$ be a sequence of subsets of $X$.
Then the
\emph{cluster} \emph{limit} of $A_n$, denoted $\mathrm{c-}\!\lim
A_n$, is defined as follows,
\begin{eqnarray*}
\mathrm{c-}\!\lim A_n=\{x\in X:\; & \exists \mbox{ sequence }x_j\in A_{n_j}
\mbox{
such that }
\\
 &
n_j\to\infty \ \mbox {and} \lim_{j\rightarrow\infty} x_j=x\}.
\end{eqnarray*}
\begin{Proposition}
\label{P3.1} $\mathrm{c-}\!\lim A_n$ is closed.
\end{Proposition}

\textsc{Proof:}\ Assume $x_j \in \mathrm{c-}\!\lim A_n$ and $x_j\rightarrow
x $. Now, $x_j\in \mathrm{c-}\!\lim A_n$ implies $\exists x_{j_k}\rightarrow
x_j$, $x_{j_k}\in A_{n(j,k)}$, where, for each $j$, $n(j,k)\rightarrow
\infty $ as $k\rightarrow \infty$.

For each $j$ there exists $K(j)$ such that
\[
d(x_{j_k},x_j)<\frac{1}{j}\mbox{ for all } k\geq K(j).
\]
Hence, we can construct a sequence $\{K(j)\}$ so that $n(j,K(j))\rightarrow%
\infty$ as $K(j)\rightarrow\infty$ and
\[
d(y_j,x_j)<\frac{1}{j},
\]
where $y_j=x_{n(j,K(j))}\in A_{n(j,K(j))}$. Hence, $\displaystyle %
\lim_{j\rightarrow\infty} y_j=\displaystyle \lim_{j\rightarrow\infty} x_j=x$%
. Thus, $x\in \mathrm{c-}\!\lim A_n$. \hfill\ $\Box$

Under fairly general circumstances, Hausdorff convergence implies cluster
limit convergence.

\begin{Proposition}
\label{P3.2}%
If $A\subset X$ is
closed and if $D(A_n,A)\rightarrow 0$, then $\mathrm{c-}\!\lim A_n=A$.
\end{Proposition}

\textsc{Proof:}\ $\mathrm{c-}\!\lim A_n \subset A$: Let $x\in \mathrm{c-}%
\!\lim A_n$. Then $\exists x_j\in A_{n_j}$ such that $x=\displaystyle \lim
x_j$. We have that $\displaystyle \sup_{y\in A_n} d(y,A)\rightarrow0$.
Hence, $d(x_j,A)\rightarrow0$. Since $z\rightarrow d(z,A)$ is continuous, $%
d(x,A)=\displaystyle \lim_{j\rightarrow\infty} d(x_j,A)=0$. Thus, since $A$
is closed, $x\in A$.

$A\subset \mathrm{c-}\!\lim A_n$: $D(A_n,A)\rightarrow0$ implies that $%
\displaystyle \sup_{y\in A} d(y,A_n)\rightarrow0$. Hence for any $x\in A$, $%
d(x,A_n)\rightarrow0$. It follows that there exists a sequence $y_n\in A_n$
such that $d(x,y_n)\rightarrow0$, i.e., $x=\displaystyle \lim_{n\rightarrow%
\infty} y_n \in \mathrm{c-}\!\lim A_n$. \hfill\ $\Box$

We will have occasion to make use of the following simple fact concerning
Hausdorff measure.

\begin{Proposition}
\label{P3.3} Let $(X,d)$ be a locally compact metric space such that all
distance balls $\{y\in X:\;d(x,y)<r\}$ are connected (e.g. a connected
Riemannian manifold). Suppose we are given subsets $A_n$, $A\subset X$ such
that $D(A_n,A)\rightarrow 0$ and $D(\partial A_n,\partial A)\rightarrow 0$.
Let $C$ be compact subset of $X$.
\begin{enumerate}
\item[{\rm (a)}]  If $C\subset \mathrm{int}\,A$ then $C\subset
\mathrm{int}A_n$ for all $%
n$ sufficiently large.
\item[{\rm (b)}]  If $C\cap \overline{A}=\emptyset $ then $C\cap
\overline{A_n}=\emptyset $ for all $n$ sufficiently large.
\end{enumerate}
\end{Proposition}

\textsc{Proof:}\ We prove only (a), as the proof of (b) is similar. To prove
(a), it is sufficient to consider the case in which $C$ is a closed metric
ball, $C=\{x: \; d(x_0,x)\leq r\}$. The proof for general $C$ follows by
taking an appropriate finite cover of $C$. Hence, in particular, we may
assume $C$ is connected.

Since $C$ is compact and contained in the interior of $A$, we have that $%
\displaystyle \inf_{x\in C} d(x,\partial A)=\delta>0$.
The convergence
$D(\partial A_n,\partial A)\rightarrow0$ then implies that $\displaystyle %
\inf_{x\in C} d(x,\partial A_n)\geq \delta/2$ for all $n$ sufficiently
large. Since we are assuming $C$ is connected, it follows that for all $n$
sufficiently large, $C\subset \mathrm{int}\,A_n$ or $C\subset X\setminus A_n$%
. If $C\subset X\setminus A_n$ for some $n$ then $\displaystyle \inf_{x\in
C} d(x,A_n)\geq \delta/2$. Since $D(A_n,A)\rightarrow0$, this cannot happen
for infinitely many $n$. Hence, $C\subset \mathrm{int}\,A_n$ for all $n$
sufficiently
large. \hfill\ $\Box$

We now pass to the space--time setting. Let $(M,g)$ be a globally hyperbolic
space--time, and $\Sigma$ be a smooth space--like Cauchy surface for
$M$.
We wish to emphasize that we shall assume global hyperbolicity throughout.
Fix a
future directed time--like vector field $T$ on $M$. Introduce a complete
Riemannian metric $h$ on $M$. Let $d$ be the associated distance function,
and let $D$ be the associated Hausdorff distance on subsets of $M$.
Normalize $T$ so that $T$ has unit length with respect to $h$.

Each integral curve of $T$ meets $\Sigma$ once and only once.
Let $\Pi
:M\rightarrow \Sigma$ be the projection onto $\Sigma$ along the integral
curves of $T$, i.e., $\Pi(p)$ is the point in $\Sigma$ where the integral
curve of $T$ through $p$ meets $\Sigma$. Let $\Psi: M\rightarrow {\mathbb R}%
\times \Sigma$ be the diffeomorphism determined by the integral curves of $T$%
, i.e., $\Psi(p)=(t,x)$, where $x=\Pi(p)$ and $t=t(p)$ is the signed $h$%
-length of the integral curve of $T$ from $x$ to $p$. When convenient, we
will surpress the map $\Psi$ and simply identify $M$ with ${\mathbb R}\times
\Sigma$. Thus, for example, by the \emph{graph} of a function $%
u:\Sigma\rightarrow{\mathbb R}$ we mean the following subset of $M$,
\[
\mbox{graph}\,u=\{(u(x),x)\in M:\; x\in\Sigma\}.
\]

We now consider the Cauchy horizons of certain subsets of $\Sigma$. For any
subset $A\subset \Sigma$, we let $\partial A=\overline A\setminus \mathrm{int%
}\,A$ denote the boundary of $A$ in $\Sigma$. Let $\hat\mathcal{X}$ be the
following collection of subsets of $\Sigma$,
\begin{eqnarray}
& \hat\mathcal{X} =\{A\subset \Sigma:%
\mbox{ $A$ is the closure of an open set in
$\Sigma$, $A$ is connected} & \nonumber{} \\
&
\mbox{and $\partial A$ is a topological submanifold
of co-dimension one in $\Sigma$}\}. \nonumber{}&
\end{eqnarray}
For any $A\in\hat\mathcal{X}$, the future Cauchy horizon
${\mathcal{H}}^+(A)$ is
an achronal Lipschitz hypersurface in $M$ such that edge ${\mathcal{H}}^+(A)=%
\mbox{edge }A =\partial A$ \cite[Propositions 5.8 and 5.11]{PenroseDiffTopo}.
Each integral curve of $T$ meets ${\mathcal{H}}^+(A)$ at most once.
Henceforth it will be convenient to define a slightly more restricted
class of sets $\mathcal{X}$ by
\begin{eqnarray}
 \mathcal{X} =\{A\in \hat\mathcal{X}: \  &\mbox{  each integral curve
of $T$  }\qquad \qquad \qquad
 \nonumber{}
\\
&
\mbox{ that meets $A$ also meets ${\mathcal{H}}^+(A)$}\}.
 \label{Xidef}
\end{eqnarray}
In this case ${\mathcal{H}}^+(A)$ may be viewed as a
graph over $A$. More precisely, there exists a locally Lipschitz function $%
u:A\rightarrow {\mathbb R}$ such that,
\[
{\mathcal{H}}^+(A)=\{(u(x),x)\in M:\; x\in A\}.
\]

For the proof of the following lemma, let $\Sigma_A$ denote the achronal
surface obtained by ``gluing together'' ${\mathcal{H}}%
^+(A)$ and $\Sigma\setminus A$ along $\partial A$. Then by setting $u(x)=0$
for all $x\in \Sigma\setminus A$, we obtain a locally Lipschitz function $%
u:\Sigma\rightarrow{\mathbb R}$ such that,
\[
\Sigma_A=\{(u(x), x)\in M:\; x\in\Sigma\}.
\]

\begin{Lemma}
\label{L3.4} Let $A_n$, $A\in \mathcal{X}$ be such that $D(A_n,A)\rightarrow
0$ and $D(\partial A_n,\partial A)\rightarrow 0$. Then for each compact set $%
C\subset A$,
\[
D({\mathcal{H}}^{+}(A_n)\cap \Pi ^{-1}(C),{\mathcal{H}}^{+}(A)\cap \Pi
^{-1}(C))\rightarrow 0.
\]
\end{Lemma}

\textsc{Proof:}\ There exist locally Lipschitz functions $%
u_n:\Sigma\rightarrow{\mathbb R}$, $n=1,2,\ldots$, and $u:\Sigma\rightarrow{%
\mathbb R}$ such that $\Sigma_{A_n}=\mbox{graph}\,u_n$ and $\Sigma_A=%
\mbox{graph}\,u$. To prove the lemma, it is sufficient to show that $%
u_n\rightarrow u$ pointwise, for then $u_n\rightarrow u$ uniformly on $C$,
and the lemma easily follows.

First suppose $x\in \Sigma\setminus A$, so that $u(x)=0$. Then, by
Proposition \ref{P3.3}, $x\in\Sigma\setminus A_n$ for all $n$ sufficiently
large, and hence $u_n(x)=0$ for all such $n$. Thus, $\displaystyle %
\lim_{n\rightarrow\infty} u_n(x)=0=u(x)$.

Now suppose $x\in\mathrm{int}\,A$, so that $(u(x),x)\in {\mathcal{H}}%
^+(A)\setminus\mbox{edge}\,A$ and $u(x)>0$. For any $\varepsilon>0$
sufficiently small, $q=(u(x)-\varepsilon,x)\in\mathrm{int}\,D^+(A)$. Hence $%
J^-(q)\cap\Sigma$ is compact and contained in $\mathrm{int}\,A$. Thus, by
Proposition \ref{P3.3}, $J^-(q)\cap\Sigma$ is contained in $A_n$ for all $n$
sufficiently large. It follows that $q\in D^+(A_n)$ for all such $n$. Hence,
$(u(x)-\varepsilon,x)$ lies to the past of $(u_n(x),x)\in {\mathcal{H}}%
^+(A_n)$ along the integral curve of $T$ through $x$. This implies that $%
u(x)-\varepsilon\leq u_n(x)$. A similar argument shows that $u_n(x)\leq
u(x)+\varepsilon$. Combining these inequalities we obtain $|u(x)-u_n(x)|\leq
\varepsilon$ for all $n$ sufficiently large. Hence, $\displaystyle %
\lim_{n\rightarrow\infty} u_n(x)=u(x)$.

Finally, suppose $x\in\partial A$. Since each 
hypersurface $\Sigma_{A_n}$ is achronal, there exists a neighbourhood $W$ of
$x$ in $\Sigma$ and a uniform Lipschitz constant $K$ such that for all $n$
and all $x_1$, $x_2\in W$,
\begin{equation}  \label{3.1}
|u_n(x_2)-u_n(x_1)|\leq K\rho(x_1,x_2),
\end{equation}
\noindent where $\rho$ is the distance function in $\Sigma$ determined by
the induced metric on $\Sigma$. The condition, $D(\partial A_n,\partial
A)\rightarrow0$ implies that there exists a sequence of points $%
y_n\in\partial A_n$ such that $y_n\rightarrow x$. Then (\ref{3.1}) implies,
\[
|u_n(x)|=|u_n(x)-u_n(y_n)|\leq K\rho(y_n,x).
\]
Since $\rho(y_n,x)\rightarrow0$, it follows that $\displaystyle %
\lim_{n\rightarrow\infty} u_n(x)=0=u(x)$. \hfill\ $\Box$

\emph{Remarks}. We note that every subset $B$ of ${\mathcal{H}}^+(A)$
is of the form ${\mathcal{H}}^+(A)\cap\Pi^{-1}(C)$, where $C=\Pi(B)$. In
particular, if $B$ is compact, $C$ is compact. We note also that Lemma \ref
{L3.4} remains valid if $C$ is merely assumed to be a subset of $A$ with
compact closure.

The following corollary is an immediate consequence of Proposition \ref{P3.2}
and Lemma \ref{L3.4}.

\begin{Corollary}
\label{C3.5} Consider $A_n$, $A\in \mathcal{X}$. Let $C\subset A$ be
compact. If $D(A_n,A)\rightarrow 0$ and $D(\partial A_n,\partial
A)\rightarrow 0$ then
\[
\mathrm{c-}\!\lim [{\mathcal{H}}^{+}(A_n)\cap \Pi ^{-1}(C)]={\mathcal{H}}%
^{+}(A)\cap \Pi ^{-1}(C).
\]
\end{Corollary}

The next lemma to be presented, which we refer to as the localization lemma,
shows that, under suitable circumstances, ``localized'' changes in the
boundary $\partial A$ produce only ``localized'' changes in ${\mathcal{H}}%
^+(A)$. The proof is a consequence of the following simple fact which
will arise in other
situations, as well. We use the convention that%
$\mathcal{D}^+(A)$, for an achronal
set $A$, is the collection of all points $p$ such that every past
inextendible \emph{time--like} curve through $p$ intersects $A$.

\begin{Proposition}
\label{P3.6} Let $A$ be a closed achronal subset of a space--time $M$. Let $%
\eta $ be a future directed causal curve from $p\in \mathrm{edge}\,A$ to $%
q\in \mathcal{D}^{+}(A)\setminus \mathrm{edge}\,A$. Then $\eta $ is
contained in ${\mathcal{H}}^{+}(A)$ and hence is a null geodesic generator
(or segment thereof) of ${\mathcal{H}}^{+}(A)$.
\end{Proposition}

\textsc{Proof:}\ $q\in \mathcal{D}^+ (A)$ forces $\eta$ to be contained in $%
\mathcal{D}^+ (A)$. If $\eta$ were to enter the interior of $\mathcal{D}^+
(A)$ then $p$ would have to be a point in $A\setminus \mathrm{edge}\,A$
(see, e.g. \cite[Prop. 5.16, p. 45]{PenroseDiffTopo}). It follows that $%
\eta\subset {\mathcal{H}}^+(A)$. Moreover, to avoid an achronality violation
of ${\mathcal{H}}^+(A)$, $\eta$ must be a null geodesic segment. \hfill\ $%
\Box$

\begin{Lemma}[The Localization Lemma]
\label{L3.7} Consider $A$, $\hat{A}\in \mathcal{X}$, with $A\subset \hat{A}$%
.  For any subset $\partial_0 A$ of $\partial A$, let
\begin{eqnarray*}
{\mathcal{H}}_0^{+}(A)=\{x\in {\mathcal{H}}^{+}(A):\;x%
\mbox{ lies on a null geodesic
  generator} \\
\mbox{ with past end point on }\partial_0 A\}.
\end{eqnarray*}
If $\partial _0A \subset \partial \hat{A}$ then ${\mathcal{H}}%
_0^{+}(A)\subset {\mathcal{H}}^{+}(\hat{A})$.
\end{Lemma}

\textsc{Proof:}\ Let $y\in {\mathcal{H}}^+_0(A)$. If $y\in \partial_0 A$,
there is nothing to show. Hence, we may assume $y\in {\mathcal{H}}%
^+_0(A)\setminus\partial A$. Let $\eta$ be a null geodesic generator of ${%
\mathcal{H}}^+(A)$ from $x\in\partial_0 A$ to $y$. Since $A\subset \hat A$, $%
\mathcal{D}^+ (A)\subset \mathcal{D}^+ (\hat A)$. Hence, $\eta$ is a null
geodesic segment from $x\in\partial\hat A=\mathrm{edge}\, \hat A$ to $y\in
\mathcal{D}^+ (\hat A)$. By Proposition \ref{P3.6}, $\eta$ is a null
geodesic generator of ${\mathcal{H}}^+(\hat A)$, and hence $y\in {\mathcal{H}%
}^+(\hat A)$. \hfill\ $\Box$

\section{Semi-tangents}

\label{S4}

We introduce the notion of semi-tangents. Let $A\in\mathcal{X}$. A
semi-tangent of ${\mathcal{H}}^+(A)$ is a vector $X\in TM$ based at a point $%
p\in {\mathcal{H}}^+(A)\setminus\partial A$, which is tangent to a null
generator $\eta$ of ${\mathcal{H}}^+(A)$, is past-directed, and is
normalized so that
$h(X,X)=1$\footnote{%
A more elegant and essentially equivalent approach would be \emph{not} to
impose the condition $h(X,X)=1$. That would avoid the usage of an auxiliary
Riemannian metric $h$. The current definition is more convenient for
technical reasons, as it fixes the parametrisation of the null geodesics
under consideration.}.

Let ${\Na }_p(A)$ denote the collection of semi-tangents of
${\mathcal{H}}^+(A)$ based at $p$,
and let $\Na (A)$ denote the collection of all semi-tangents of
${\mathcal{H}}^+(A)$.
 Note that $\Na (A)\subset \UM $, where $\UM $ is the
unit tangent bundle of $M$ with respect to the Riemannian metric $h$. If we
let $M_A=M\setminus \partial A$, then $M_A$ is an open submanifold of $M$
and $\UM _A$ is an open sub--bundle of $\UM$. Since, in
the definition of $\Na (A)$, vectors based at $\partial A$ are excluded, we
have that $\Na (A)\subset \UM_A$.

\begin{Lemma}
\label{L4.1} $N^{+}(A)$ is closed in $\UM_A$.
\end{Lemma}

\textsc{Proof:}\ Let $X_n\in \Na (A)$ such that $X_n\rightarrow X$ in $%
\UM_A$.
Let $X_n$ be based at the point $p_n\in {\mathcal{H}}%
^+(A) $ and $X$ be based at $p$. Since $p_n\rightarrow p$, $p\in {\mathcal{H}%
}^+(A)\setminus \partial A$. Let $\eta_n$ be the past directed null
generator of ${\mathcal{H}}^+(A)$ starting at $p_n$ with initial tangent $X_n
$. Let $q_n\in \partial A$ be the past end point of $\eta_n$. The $\eta_n$'s
converge, as geodesics, to a past directed null geodesic $\eta$ which starts
at $p$ with initial tangent $X$. $\eta$ will meet $\Sigma$ at a point $q\in
\mathcal{D}^+ (A)$. Convergence properties of geodesics guarantee that $%
q_n\rightarrow q$. Hence, $q\in\partial A$. Thus $\eta$ is a null geodesic
segment from $p\in {\mathcal{H}}^+(A)$ to $q\in\partial A$. By Proposition
\ref{P3.6}, this forces $\eta$ to be a null geodesic generator of ${\mathcal{%
H}}^+(A)$. Thus, $X\in \Na (A)$. \hfill\ $\Box$

We now establish some connections between the distribution of semi-tangents
and the regularity of Cauchy horizons. Lemma \ref{L4.1} may be used to show
that the distribution of semi-tangents is continuous at \emph{interior}
points (non-end points) of null generators. For $A\in\mathcal{X}$, consider
the set,
\[
S=\{x\in {\mathcal{H}}^+(A):\; x%
\mbox{ is an interior point of a null generator of
}{\mathcal{H}}^+(A)\},
\]
endowed with the subspace topology. We note the following well known result:

\begin{Lemma}
\label{L4.1.1} At each point $p\in S$ there exists a \emph{unique}
semi-tangent $X(p)\in N^{+}(A)$.
\end{Lemma}

\textsc{Proof:}\ Indeed, if this were not the case, there would be one null
generator meeting another at an interior point. Then, traveling around the
corner at which they meet would produce an achronality violation. \hfill\ $%
\Box$

Lemma \ref{L4.1.1} shows that there is a well-defined map $\Psi
:S\rightarrow N^{+}(A)$ defined by $\Psi (p)=X(p)$. If we endow $N^{+}(A)$ $%
\subset TM$ with the subspace topology, we obtain:%

\begin{Proposition}
\label{P4.2} $\Psi :S\rightarrow N^{+}(A)$ is continuous.
\end{Proposition}

\textsc{Proof:}\ Given $p\in S$, consider any sequence $\{p_n\}\subset S$
such that $p_n\rightarrow p$. Let%
$X_n=\Psi (p_n)$ and $X=\Psi (p)$. We want to show
that $X_n\rightarrow X$. Indeed, since the $X_n$'s are $h$-unit vectors,
there exists a subsequence $X_n$, that converges to an $h$-unit vector $Y$.
By Lemma \ref{L4.1}, $Y$ is a semi-tangent based at the interior point $p$.
But since semi-tangents at interior points are unique, we must have $Y=X$,
which had to be established. \hfill\ $\Box $

The interest in the distribution of semi-tangents lies in the fact,
that it
encodes most --- if not all --- information about differentiability of ${%
\mathcal{H}}^{+}(A)$. Indeed, at any point of ${\mathcal{H}}^{+}(A)$ where
there is more than one semi-tangent, the Cauchy horizon must fail to be
differentiable.

\begin{Proposition}
\label{P4.4} Consider the Cauchy horizon ${\mathcal{H}}^{+}(A)$, $A\in
\mathcal{X}$. If there exist two distinct semi-tangents at $p\in {\mathcal{H}%
}^{+}(A)\setminus \partial A$ then ${\mathcal{H}}^{+}(A)$ is not
differentiable at $p$.
\end{Proposition}

The heuristic idea is clear. If ${\mathcal{H}}^+(A)$ were differentiable at $%
p$ it would have a tangent hyper-plane $\Pi_p$ at $p$. Since ${\mathcal{H}}%
^+(A)$ is achronal and generated by null geodesics, $\Pi_p$ should be a null
hyper-plane. On the other hand, $\Pi_p$ should contain the two distinct
semi-tangents at $p$, forcing $\Pi_p$ to be a time--like hyper plane. The
proof presented below carries this argument out rigorously.

\textsc{Proof:}\ Suppose to the contrary that ${\mathcal{H}}^{+}(A)$ is
differentiable at $p$. Introduce geodesic normal coordinates $%
(x^0,x^1,\ldots ,x^n)=(x^0,\vec{x})$ centered at $p$, defined in a
neighborhood $\mathcal{U}$ of $p$ such that $\displaystyle\left. \frac \partial
{\partial x^0}\right| _p$ is future pointing time--like. Choose $\mathcal{U}$
sufficiently small so that $\displaystyle\frac \partial {\partial x^0}$ is
time--like on $\mathcal{U}$. Any geodesic $\gamma $ through $p$, when
expressed in these coordinates takes the form,
\[
\gamma :\quad x^\mu (s)=\lambda ^\mu s,\quad \lambda ^\mu \in {\mathbb R}%
,\quad \mu =0,\ldots ,n.
\]
$\gamma $ is time--like, space--like,
null if and only if $-|\lambda ^0|^2+|%
\vec{\lambda}|^2<0$, $>0$, $=0$.
It follows that the future (past) null cone at $p$ is
described by the graph $x^0=|\vec{x}|$ ($x^0=-|\vec{x}|$).

Since ${\mathcal{H}}^{+}(A)$ is achronal, it can be expressed in $\mathcal{U}
$ as a graph over the slice $\mathcal{V}=\{x^0=0\}$,
\[
{\mathcal{H}}^{+}(A)\cap \mathcal U=\{\quad x^0=f(\vec{x}),\quad \vec{x}\in
\mathcal{%
V\}\;}.
\]
${\mathcal{H}}^{+}(A)$ is differentiable at $p$ if and only if the function $%
f(\vec{x})$ is differentiable at $\vec{0}$, i.e. iff there exists a vector $%
\vec{C}=(C^1,\ldots ,C^n)$ such that,
\begin{equation}
\lim_{\vec{x}\rightarrow 0}\frac{f(\vec{x})-\vec{C}\cdot \vec{x}}{|\vec{x}|}%
=0\;.  \label{4.1}
\end{equation}

\emph{Claim}. $|\vec C|\leq 1$.

Otherwise, as we show, the achronality of ${\mathcal{H}}^{+}(A)$ is
violated. Suppose, then, that $|\vec{C}|>1$. By (\ref{4.1}), with $\vec{x}=t%
\vec{C}$, we have that for any $\varepsilon >0$ $\exists \;t_0>0$ such that
for all $t\in (0,t_0)$,
\[
\frac{|f(t\vec{C})-|\vec{C}|^2t|}{|\vec{C}|t}<\varepsilon \;,
\]
which implies
\[
f(t\vec{C})>t|\vec{C}|(|\vec{C}|-\varepsilon )>t|\vec{C}|,
\]
provided $\varepsilon $ is sufficiently small. But this implies that ${%
\mathcal{H}}^{+}(A)$ enters into the future null cone at $p$, which violates
the achronality of ${\mathcal{H}}^{+}(A)$.

Let $\overline{A}=(A^0,\vec{A})$ and $\overline{B}=(B^0,\vec{B})$ be any two
semi-tangents at $p$,
scaled so that $|\vec{A}| = |\vec{B}| = 1$.
Since $\overline{A}$ and $\overline{B}$ are past directed null, $A^0=B^0 = -1$.
Then $s\rightarrow (sA^0,s\vec{A})$ and $s\rightarrow (sB^0,s%
\vec{B})$ are past directed null geodesic generators of ${\mathcal{H}}^{+}(A)
$. It follows that $f(s\vec{A})=sA^0=-s$, and $f(s\vec{B})=-s$.
Setting $\vec{x}=s\vec{A}$ in (\ref{4.1}) gives,
\[
\lim_{s\rightarrow 0}\frac{s+s\vec{C}\cdot \vec{A}}{s}=0
\]
which implies that
\[
-1=\vec{C}\cdot \vec{A}=|\vec{C}|\,\cos \theta .
\]
Hence $\cos \theta =-\frac 1{|\vec{C}|}\leq -1$, forcing $\theta =\pi$ and
$|\vec C| =1$. Thus,
$\vec{A}= -\vec{C}$. But the same argument also shows that
$\vec{B}= -\vec{C}$. Hence, $\vec{A}=\vec{B}$, which contradicts the
assumption that there are
distinct semi-tangents at $p$.
\hfill\ $\Box $

Proposition \ref{P4.4} shows what happens at those points at which $N^{+}(A)$
consists of more than one element. We do not know what happens at all points
at which $N^{+}(A)$ contains only one vector (and it would be of
interest to settle that question). However, when those
points are interior points of generators,
we have the following result.

\begin{Proposition}
\label{P4.3} At interior points of null generators the Cauchy horizon ${%
\mathcal{H}}^{+}(A)$ is differentiable.
\end{Proposition}

Related results concerning the regularity
of Cauchy horizons (or, more generally, achronal sets) at interior
points of generators have been  obtained by I. Major \cite{Major} and K. Newman
\cite{Newman:private}.

\textsc{Proof:}\ Let $\gamma $ be a null generator of
${\mathcal{H}}^{+}(A)$ and let $p$ be any of its interior points. Let
$\mathcal{U}$, $x^\mu $, $f$, etc., be as in the proof of Proposition
\ref{P4.4}. Note that any point $p=(t,\vec x) \in \mathcal{U}\cap
\mathcal{D}^{+}(A)$ satisfies $t\le f(\vec{x})$. Similarly if
$p=(t,\vec x) \in \mathcal{U}$ is not in $ \mathcal{D}^{+}(A)$ then
$t\ge f(\vec{x})$.  We can find $q^{+}\in \gamma \cap J^{+}(p)$ such
that, passing to a subset of $\mathcal{U}$ if necessary,
$J^{-}(q^{+})$ is a smooth submanifold of $\mathcal{U}$, given as a
graph $x^0=f_{+}(\vec{x})$. Similarly we can find $q^{-}\in \gamma
\cap J^{-}(p)$ such that $J^{+}(q^{-})$ is a smooth submanifold of
$\mathcal{U}$, given as a graph $%
x^0=f_{-}(\vec{x})$. As every point to the past of $q^{+}$ lies inside of $%
\mathcal{D}^{+}(A)$ we have
\[
f_{+}(\vec{x})\leq f(\vec{x}).
\]
Similarly every point to the future of $q^{-}$ lies outside of $\mathcal{D}%
^{+}(A)$, so that we have
\[
f(\vec{x})\leq f_{-}(\vec{x}).
\]
At the origin we have $f(0)=f_{+}(0)=f_{-}(0)$. Moreover
$J^{+}(q^{-})$ and $ J^{-}(q^{+})$ are tangent there, so that there
exists a vector $\vec{C}$ such that $\vec{C}=df_{+}(0)=df_{-}(0).$ It
follows that
\[
\vec{C}\vec{x}+o(|\vec{x}|)=f_{+}(\vec{x})\leq f(\vec{x})\leq f_{-}(\vec{x})=%
\vec{C}\vec{x}+o(|\vec{x}|),
\]
which establishes (\ref{4.1}) and shows differentiability of ${\mathcal{H}}%
^{+}(A)$ at $p$.\hfill$\Box$

\section{Convergence in the Distribution of Semi--tangents}

\label{S5}
In this section we establish some further properties of the
distribution of semi--tangents. We retain all
the notational conventions from previous sections.
In particular, here as elsewhere, global hyperbolicity of the
space--time is assumed. Let $\hat{\Pi}=\Pi\circ
P:\UM\rightarrow\Sigma$, where $P:\UM\rightarrow M$
is the natural
projection map. Thus, if $C\subset\Sigma$, $\hat{\Pi}^{-1}(C)$ is the set of
all $h$-unit vectors over $\Pi^{-1}(C)\subset M$.

\begin{Lemma}
\label{L5.1}
Consider $A_n$, $A\in \mathcal{X}$ such that $D(A_n,A)$ and
$D(\partial A_n,\partial A)$ tend to zero as $n$ tends to infinity.
\begin{itemize}
\item[(a)]  If $C$ is compact and $C\subset \mathrm{int}\,A$ then
\begin{equation}
\mathrm{c-}\!\lim [N^{+}(A_n)\cap \hat{\Pi}^{-1}(C)]\subset N^{+}(A)\cap
\hat{\Pi}^{-1}(C).  \label{5.1}
\end{equation}
\item[(b)]  If $C$ is open (as a subset of $\Sigma$) and $C\subset A$ then,

\begin{equation}
N^{+}(A)\cap \hat{\Pi}^{-1}(C)\subset \mathrm{c-}\!\lim [N^{+}(A_n)\cap \hat{%
\Pi}^{-1}(C)].  \label{5.2}
\end{equation}
\end{itemize}
\end{Lemma}

\emph{Remark}. Following the proof of Lemma \ref{L5.1} we present an
example which shows that the inclusion in (\ref{5.1}) (resp. (\ref{5.2}))
can be strict (even if, in (\ref{5.2}), one takes the \emph{closure} of
$\Na (A)\cap\hat{\Pi}^{-1}(C)$).

\textsc{Proof:}\ (a): Let $X_j\in \Na (A_{n_j})\cap\hat{\Pi}^{-1}(C)$ such
that $X_j\rightarrow X$ in $\UM$. Since $\hat{\Pi}^{-1}(C)$ is
closed, $X\in \hat{\Pi}^{-1}(C)$.
Let $X_j$ be based at $p_j\in {\mathcal{H}}%
^+(A_{n_j})$, and let $X$ be based at $p$. Let $\eta_j$ be the past directed
null geodesic generator of ${\mathcal{H}}^+(A_{n_j})$ with initial point $%
p_j $ and initial tangent $X_j$, and let $q_j\in \partial A_{n_j}$ be its
past end point. Let $\eta$ be the past directed null geodesic with initial
point $p$ and initial tangent $X$. Since $p_j\rightarrow p$, Corollary \ref
{C3.5} implies that $p\in {\mathcal{H}}^+(A)\setminus \partial A$. Then $\eta
$ meets $\Sigma$ at some point $q\in A$.
Since $\eta_j\rightarrow\eta$ in the sense of geodesics it follows that $%
q_j\rightarrow q$. The assumption $D(\partial A_n,\partial A)\rightarrow0$
implies that $q\in\partial A$. Hence, by Proposition \ref{P3.6}, $\eta$ is a
null geodesic generator of ${\mathcal{H}}^+(A)$, and thus $X\in \Na (A)\cap
\hat{\Pi}^{-1}(C)$.

(b): Consider $X\in \Na (A)\cap \hat{\Pi}^{-1}(C)$ based at $p\in {\mathcal{H}%
}^+(A)\cap\Pi^{-1}(C)$ (in particular, $p\notin\partial A$). Let $\eta$
be the past directed null generator of ${\mathcal{H}}^+(A)$ starting at $p$
with initial tangent $X$, and let $q\in\partial A$ be the past end point of $%
\eta$. Let $y\in {\mathcal{H}}^+(A)\cap\Pi^{-1}(C)$ be an interior point of $%
\eta$. Since $C$ is open, $y$ can be chosen arbitrarily close to $p$. By
Corollary \ref{C3.5} (with $C=\Pi(y)$) there exists $y_j\in {\mathcal{H}}%
^+(A_{n_j})\cap\Pi^{-1}(C)$ such that $y_j\rightarrow y$. Let $\mu_j$ be a
past directed null generator of ${\mathcal{H}}^+(A_{n_j})$ starting at $y_j$
with initial tangent $Y_j\in \Na (A_{n_j})$. Assume $\mu_j$ meets $\partial
A_{n_j}$ at $z_j$. For any $p^{\prime}\in I^+(p)$, the sequence $\{z_j\}$
eventually enters the set $J^-(p^{\prime})\cap\Sigma$. Since, by the global
hyperbolicity of $M$, this set is compact, $\{z_j\}$ has an accumulation
point $z$. Proposition \ref{P3.2} implies that $z\in\partial A$. Thus, by
passing to a subsequence if necessary, we may assume the $\mu_j$ converge to
a null geodesic $\mu$ with future end point $y$ and past end point $z$.
Proposition \ref{P3.6} implies that $\mu$ is a null geodesic generator of ${%
  \mathcal{H}}^+(A)$. In order to avoid there being two distinct
semi--tangents at the interior point $y$, $\mu$ must coincide with
$\eta$, {\em cf.\,} Lemma \ref{L4.1.1}. Hence, the
tangent of $\eta$ at $y$ coincides with the tangent of $\mu$ at $y\in\mathrm{%
c-}\!\lim \Na (A_n)$. Since $y$ can be chosen arbitrarily close to $p$,
Proposition \ref{P3.1} implies that $X\in\mathrm{c-}\!\lim \Na (A_n)$. \hfill%
\ $\Box$

\begin{example}
\label{e5.2}%
Consider Minkowski $2$-space, $M={\mathbb R}^{1,1}$, with metric
$\mathrm{d}s^2=\mathrm{d}x^2-\mathrm{d}y^2$. Let $\Sigma $ be the
slice $y=0$. Set, $A=\{(x,0):\;x\in (-2,2)\}$, $A_n=\{(x,0):\;x\in
(-2-\frac 1n,2-\frac 1n)\}$.
If $C=\{(x,0):\;x\in [0,1]\}$, the
inclusion (\ref{5.1}) is strict. If $C=\{(x,0):\;x\in (-1,0)\}$, the
inclusion (\ref {5.2}) is strict (even if one takes the closure of
$N^{+}(A)\cap \hat{\Pi} ^{-1}(C)$).
\end{example}

We now wish to consider the Hausdorff convergence of certain subsets of the
distribution of semi--tangents. For this purpose, introduce a complete
Riemannian
metric on the $h$-unit tangent bundle $\UM$, and let $\rho$
denote the associated distance function. With respect to subsets $V$, $%
W\subset \UM$, let $\mathcal{D}(V,W)$ denote the Hausdorff
distance between $V$ and $W$ with respect to the distance function $\rho$.

Given $A_n$, $A\in\mathcal{X}$ such that $D(A_n,A)\rightarrow0$ and $%
D(\partial A_n, \partial A)\rightarrow0$, one might have hoped that, for
reasonable subsets $C\subset \mathrm{int}A$ (e.g. $C$ compact or open with
compact closure), $\Na (A_n)\cap \hat{\Pi}^{-1}(C)$ Hausdorff converges to $%
\Na (A)\cap \hat{\Pi}^{-1}(C)$, i.e., $\mathcal{D} (\Na (A_n)\cap \hat{\Pi}%
^{-1}(C),\Na (A)\cap \hat{\Pi}^{-1}(C))\rightarrow0$ as $n\rightarrow\infty$.
But Example \ref{e5.2} shows that this will not be the case in general.
However, it is possible to establish a slightly weaker result that is
sufficient for our purposes.

\begin{Theorem}
\label{T5.3} Consider $A_n$, $A\in \mathcal{X}$ such that $D(A_n,A)$ and
$D(\partial A_n,\partial A)$ tend to zero as $n$ tends to infinity.
\begin{itemize}
\item[(a)]  If $C$ is compact and $C\subset \mathrm{int}A$ then,
\[
\sup_{X\in N^{+}(A_n)\cap \hat{\Pi}^{-1}(C)}\rho (X,N^{+}(A)\cap \hat{\Pi}%
^{-1}(C))\rightarrow 0\quad \mathrm{as}\quad n\rightarrow \infty .
\]
\item[(b)]  If $C$ is open with compact closure and $\overline{C}\subset
\mathrm{int}A$ then,
\[
\sup_{Y\in N^{+}(A)\cap \hat{\Pi}^{-1}(C)}\rho (Y,N^{+}(A_n)\cap \hat{\Pi}%
^{-1}(C))\rightarrow 0\quad \mathrm{as}\quad n\rightarrow \infty .
\]
\end{itemize}
\end{Theorem}

Part (a) is a consequence of Lemma \ref{L4.1}, part (a) of Lemma
\ref{L5.1} and the following simple fact.

\begin{Lemma}
\label{L5.4} Let $(X,d)$ be a metric space. Consider subsets $A_n$, $%
A\subset X$. Assume the $A_n$'s are closed. Suppose there exists a compact
set $B\subset X$ such that $A_n\subset B$ for all $n$. If $\mathrm{c-}\!\lim
A_n\subset A$ then,
\[
\sup_{x\in A_n}d(x,A)\rightarrow 0\quad \mathrm{as}\quad n\rightarrow \infty
.
\]
\end{Lemma}

\textsc{Proof:}\ of Lemma \ref{L5.4}. Note that the $A_n$'s are compact. If
the conclusion does not hold then there exists $\varepsilon>0$ and a
subsequence $A_{n_k}$ such that,
\[
\sup_{x\in A_{n_k}} d(x,A)\geq\varepsilon\quad \mbox{for all}\quad k.
\]
By the compactness of $A_{n_k}$, $\exists x_k\in A_{n_k}$ such that $%
d(x_k,A)\geq \varepsilon$. Since $\{x_k\}\subset B$, by passing to
subsequence if necessary, we may assume $x_k\rightarrow y$. Since $\mathrm{c-%
}\!\lim A_n\subset A$, $y\in A$. On the other hand,
\[
d(y,A)=\lim_{k\rightarrow\infty} d(x_k,A)\geq\varepsilon,
\]
and hence $y\notin A$, a contradiction. \hfill\ $\Box$

\textsc{Proof:}\ of part (a) of Theorem \ref{T5.3}. The aim is to apply
Lemma \ref{L5.4}.
The set $B:= {\mathcal{H}}^+(A)\cap\Pi^{-1}(C)$ is compact and does not meet
$\partial A$. Note that $\Na (A)\cap\hat{\Pi}^{-1}(C)=\Na (A)\cap P^{-1}(B)$,
where, recall, $P:\UM\rightarrow M$ is the natural projection
map. Since $P^{-1}(B)$ is compact and $\Na (A)$ is closed in $\mathcal{U}%
\!M_A $ (recall that $M_A=M\setminus\partial A$) by Lemma \ref{L4.1},
 $\Na (A)\cap\hat{\Pi}^{-1}(C)$ is compact. Similarly, using Proposition
\ref{P3.3}, for sufficiently large $n$, $\Na (A_n)\cap\hat{\Pi}^{-1}(C)$ is
compact. Since, by Lemma \ref{L3.4}, $D({\mathcal{H}}^+(A_n)\cap\Pi^{-1}(C),
{\mathcal{H}}^+(A)\cap\Pi^{-1}(C))\rightarrow0$, it follows that
${\mathcal{H}}^+(A_n)\cap\hat{\Pi}^{-1}(C)$
($n$ sufficiently large) is contained in a compact set $K\subset M$. (One
may take $K$ to be the closure of an $\varepsilon$-neighborhood of ${%
\mathcal{H}}^+(A)\cap\hat{\Pi}^{-1}(C)$). It follows that
$\Na (A_n)\cap\hat{\Pi}^{-1}(C)$ (for $n$ sufficiently large)
is contained in the compact set $P^{-1}(K)\subset \UM$. Also,
part (a) of Lemma \ref{L5.1} implies that $\mathrm{c-}\!\lim
[\Na (A_n)\cap\hat{\Pi}%
^{-1}(C)] \subset \Na (A)\cap\hat{\Pi}^{-1}(C)$. Part (a) of Theorem
\ref{T5.3} now
follows directly from Lemma \ref{L5.4}.

\textsc{Proof:}\ of part (b) of Theorem \ref{T5.3}. Suppose our claim
does not hold. Then
there exists $\varepsilon>0$ and a sequence $n_j$ such that,
\begin{equation}  \label{5.3}
\sup_{Y\in \Na (A)\cap\hat{\Pi}^{-1}(C)} \rho(Y,\Na (A_{n_j})\cap\hat{\Pi}%
^{-1}(C)) \geq 4\varepsilon.
\end{equation}
Let $Y_j\in {\cal U}_{p_j}M$ be such that the sup in (\ref{5.3}) is
obtained at $Y=Y_j$. By the compactness of
$\overline{\Na (A)\cap\hat{\Pi}^{-1}(C)}$, there exists $Y\in
\overline{\Na (A)\cap\hat{\Pi}^{-1}(C)}$ such that, passing to a
subsequence if necessary, we have $Y_j\rightarrow Y$. It follows that
for $j$ large enough we shall have,
\begin{equation}  \label{5.4}
\rho(Y,\Na (A_{n_j})\cap\hat{\Pi}^{-1}(C))\geq 2\varepsilon.
\end{equation}

Since every neighborhood of $Y$ meets $\Na (A)\cap\hat{\Pi}^{-1}(C)$, we can
choose $\tilde Y\in \Na (A)\cap\Pi^{-1}(C)$ such that $\rho(Y,\tilde
Y)\leq\varepsilon/2$. Since $C$ is open, there will exist a semi-tangent $%
\hat Y\in \Na (A)\cap\hat{\Pi}^{-1}(C)$ at an interior point $\hat p$ of the
null geodesic generator of ${\mathcal{H}}^+(A)$ determined by $\tilde Y$,
such that $\rho(\hat Y,\tilde Y)\leq\varepsilon/2$. It follows that for all $%
j$ large enough,
\begin{equation}  \label{5.5}
\rho(\hat Y,\Na (A_{n_j})\cap\hat{\Pi}^{-1}(C)\geq\varepsilon.
\end{equation}

By Corollary \ref{C3.5} there exists a sequence $q_j\in {\mathcal{H}}%
^+(A_{n_j})\cap\hat{\Pi}^{-1}(C)$ such that $q_j\rightarrow \hat p$.
(\ref {5.5}) implies there exists a sequence $Z_j\in \Na (A_{n_j})\cap
{\cal U}_{p_j}M$
such that $\rho(\hat Y,Z_j) \ge \epsilon$ for all $j$ sufficiently large.
By passing to a subsequence if necessary, we
have that $Z_j\rightarrow Z\in \overline{U}_{\hat p}M$.
Proposition \ref{P3.1} and part (b) of Lemma \ref{L5.1} imply that $Z\in
\Na (A)$. Clearly $Z\neq \hat Y $ by (\ref{5.5}), which contradicts the
fact that there can be only one
semi-tangent at an interior point of a null geodesic generator. \hfill\ $%
\Box $

\section{The proof of Theorem {\protect \ref{T1}}}

\label{SP}

Let us start by giving an outline of the construction of the set $K$, the
existence of which is asserted in Theorem \ref{T1}. The idea is to construct
a sequence of sets $K_i$, converging in a precise sense to the desired set $%
K $, such that

\begin{enumerate}
\item  $K_0={\mathbb R}^3\setminus B(0,1)$, where $B(x_0,r)$ denotes an open
ball of radius $r$ centered at $x_0$.

\item  $\partial K_{i+1}$ is obtained by adding a certain number of
``ripples'' to $\partial K_i$. The process of adding ripples consists of
adding a certain number of circular arcs to $\partial K_i$. It follows that $%
\partial K_i$ is the union of a finite number of circular arcs.

\item The horizon ${\mathcal{H}}^{+}( K_i)$ has several ``creases'' --
  curves along which ${\mathcal{H}}^{+}(K_i)$ is not differentiable.
  The addition of ripples to $\partial K_i$ does not affect the
  existence of the ``old'' creases, and leads to the occurrence of new
  ones. The construction guarantees that the set of points which lie
  on creases of ${\mathcal{H}}^{+}(K)$ is dense.

\item The boundaries of the $K_i$'s are radial graphs, that is,
  $\partial K_i$ is described by an equation $r=f_i(\theta )$, $\theta
  \in [0,2\pi ]$, for some Lipschitz continuous function $f_i(\theta
  )$. The construction guarantees that $\partial K$ will also have
  that property, hence will be a Lipschitz continuous topological
  manifold.
\end{enumerate}

Before we proceed further, let us recall some elementary facts from
Lorentzian geometry. Consider a globally hyperbolic space--time $(M, g)$ with
Cauchy surface $\Sigma$. Let $\Omega$ be an open subset of $\Sigma$, then ${%
\mathcal{H}}^+(\overline{\Omega})$ is a union of null geodesic segments $%
\gamma$ (\cite[Theorem 5.12]{PenroseDiffTopo}, \cite[Prop. 53, p.
430]{BONeill}),
called generators of ${\mathcal{H}}^+(\overline{\Omega})$. Each such
generator has
a past end point $p\in\partial \Omega=\overline{\Omega}\setminus\Omega$, and
either no future end point, or a future end point
$q\in{\mathcal{H}}^+(\overline{\Omega})$.
If $\partial \Omega$ is $C^2$  near $p$, then
$\gamma$ is the unique null geodesic orthogonal to $%
\partial\Omega$ pointing towards $\Omega$ \cite[Lemma 50, p. 298]{BONeill}.

Let us now pass to the description of the process that we call ``adding a
ripple''. More precisely, we will be adding a ripple at a
 point $y_0$ lying in the ``middle'' of a circular arc $\mathcal{A}$.
Consider thus an arc $\mathcal{A}$ of
a circle
$S(x_0,r_0)$ of radius
$r_0$
 centered at $x_0$, thus $\mathcal{A}$ is given by the equation $\mathcal{A}%
 =\{\novec{x}:\; \novec{x}-\novec{x}_0=r_0\mathrm{e}^{i\varphi}, \varphi\in[%
 \varphi_-,\varphi_+] \}$.
 We will be interested in the structure of a
 Cauchy horizon of a set which has first $\mathcal{A}$, and then
 ``$\mathcal{A}$ plus a ripple'' as part of its boundary. Now for any
 set $\Omega$ and any conformal isometry $\Psi$ we have
 ${\mathcal{H}}^+(\Psi(\Omega)) =\Psi({\mathcal{H}}^+(\Omega))$.
We can
 always find a conformal isometry $\Psi$ of the three-dimensional
 Minkowski
 space--time ${\mathbb R}^{2,1}$ so that $x_0=0$,
$r_0=1$, and $\varphi\in[%
 \frac{\pi}2-\varphi_0,\frac{\pi}2+\varphi_0]$ on $\Psi(\mathcal{A})$, and
it follows that
 for our purposes it is sufficient to consider this case.
Then the addition
 of a ripple to $\Psi(\mathcal{A})$ at $y_0=\mathrm{e}^{i\frac{\pi}2}$
proceeds as  follows. Let $0<r_1<1$, $0<\varphi_1<\varphi_0$, and let
$x_1=(1-r_1)\mathrm{e}^{i(\frac{\pi}2-\varphi_1)}$,
$x_2=(1-r_1)\mathrm{e}^{i(\frac{\pi}2+\varphi_1)}$.
Consider
 the circles $S(x_1,r_1)$ and $S(x_2, r_1)$, they are tangent to $\mathcal{A}$
 at points $y_1=\mathrm{e}^{i(\frac{\pi}2-\varphi_1)}$ and $y_2=\mathrm{e}%
 ^{i(\frac{\pi}2-\varphi_2)}$. If $r_1$ is large enough compared with
$|x_1-x_2|$, $%
 r_1\geq a=\frac{|x_1-x_2|}{2}$, then $S(x_1,r_1)$ will intersect $%
 S(x_2,r_1)$,
and henceforth we will assume that this is the case.
 The process of adding a ripple to $\mathcal{A}$ is made clear by
 Figure \ref{addingaripple}, and can formally be described as follows:
 \begin{figure}[thbp]
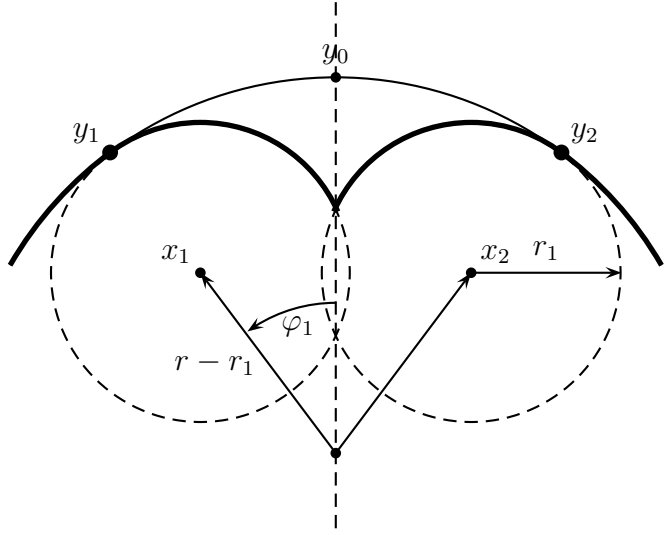

 \centering
\pspicture(-4,-1)(4,6)
\psarc(0,0){5}{30}{150}
\uput[135](-3,4){$y_1$}
\uput[45](3,4){$y_2$}
\pscircle*(-3,4){3pt}
\pscircle*(3,4){3pt}
\pscircle*(0,0){2pt}
\pscircle*(1.8,2.4){2pt}
\pscircle*(0,5){2pt}
\uput[90](0,5){$y_0$}
\uput[45](1.8,2.4){$x_2$}
\pscircle*(-1.8,2.4){2pt}
\uput[135](-1.8,2.4){$x_1$}
\psline{->}(0,0)(1.8,2.4)
\uput[180](-0.9,1.2){$r-r_1$}
\psline{<-}(3.8,2.4)(1.8,2.4)
\uput[90](2.8,2.4){$r_1$}
\psline{->}(0,0)(-1.8,2.4)
\pscircle[linestyle=dashed](-1.8,2.4){2}
\pscircle[linestyle=dashed](1.8,2.4){2}
\psarc{->}(0,0){2}{90}{126}
\uput[270](-0.5,2){$\varphi_1$}
\psline[linestyle=dashed](0,-1)(0,6)
\psarc[linewidth=2pt](-1.8,2.4){2}{24}{130}
\psarc[linewidth=2pt](1.8,2.4){2}{50}{156}
\psarc[linewidth=2pt](0,0){5}{126}{150}
\psarc[linewidth=2pt](0,0){5}{30}{54}
\endpspicture
\caption{The process of adding a ripple: the thin solid line represents the
original arc, the thick solid line corresponds to ``an arc with a ripple
added''. }
\label{addingaripple}
\end{figure}
``$\Psi(\mathcal{A} \ \mbox{with a ripple added})$'' will be the curve
obtained by
following $\Psi(\mathcal{A})$ from
$\mathrm{e}^{i(\frac{\pi}2-\varphi_0)}$ to $
\mathrm{e}^{i(\frac{\pi}2-\varphi_1)}$, then following $S(x_1,r_1)$
from $\mathrm{e}^{i(\frac{\pi}2-\varphi_1)}$ to the $y$-axis where
$S(x_1,r_1)$ intersects $S(x_2,r_1)$, then following $S(x_2,r_1)$
to $\mathrm{e}^{i(\frac{\pi}2+\varphi_1)}$%
, and finally following $\Psi(\mathcal{A})$ to its end point. It should be
seen that this construction is invariant under reflections across the $y$%
-axis.

The parameter $r_1$ in the construction will be called the radius of
the ripple. The distance $a=\frac12|\novec{x}_1-\novec{x}_2|$ will be
called the diameter of the ripple. It should be noted that
``$\mathcal{A}$ with a ripple'' can be described as a radial graph
$|\novec x-\novec x_0|=g(\varphi)$, $\varphi\in[ \varphi_-,\varphi_+]$.
$g(\varphi)$ is Lipschitz continuous and piecewise smooth. Let us for
further use note that the modulus of Lipschitz
continuity $L=L(r_1,a)$ of $g(\varphi)$ satisfies $L\leq \displaystyle %
\lim_{\varphi\rightarrow\frac{\pi}2} \left| g^{\prime}(\varphi)\right|$.
$L$ is a
strictly decreasing function 
of $a$, with $\lim_{a\to 0}L(r_1,a)=0$.

Let us now address the question, how does the addition of a ripple affect
the Cauchy horizon of $K_0={\mathbb R}^2\setminus B(0,1)$. As discussed at
the beginning of this section, generators of the horizon of any set are null
geodesics, which in our case are null straight lines in ${\mathbb R}^{2,1}$.
Every such line is determined uniquely by its projection on ${\mathbb R}^2$
under the projection map $\Pi:{\mathbb R}^{2,1}\rightarrow{\mathbb R}^2$, $%
\Pi(t,\novec x)=\novec x$, and this projection is itself a straight line
segment in ${\mathbb R}^2$. The generators emanating from points $p$
lying on a smooth piece of the boundary of the set under consideration
are thus straight lines orthogonal to the boundary. It follows that
the generators of ${\mathcal{ H}}^+(K_0)$ project down to the radial
lines $\{x=r\mathrm{e} ^{i\varphi},r\in[1,\infty)\}$.

\begin{figure}[thbp]
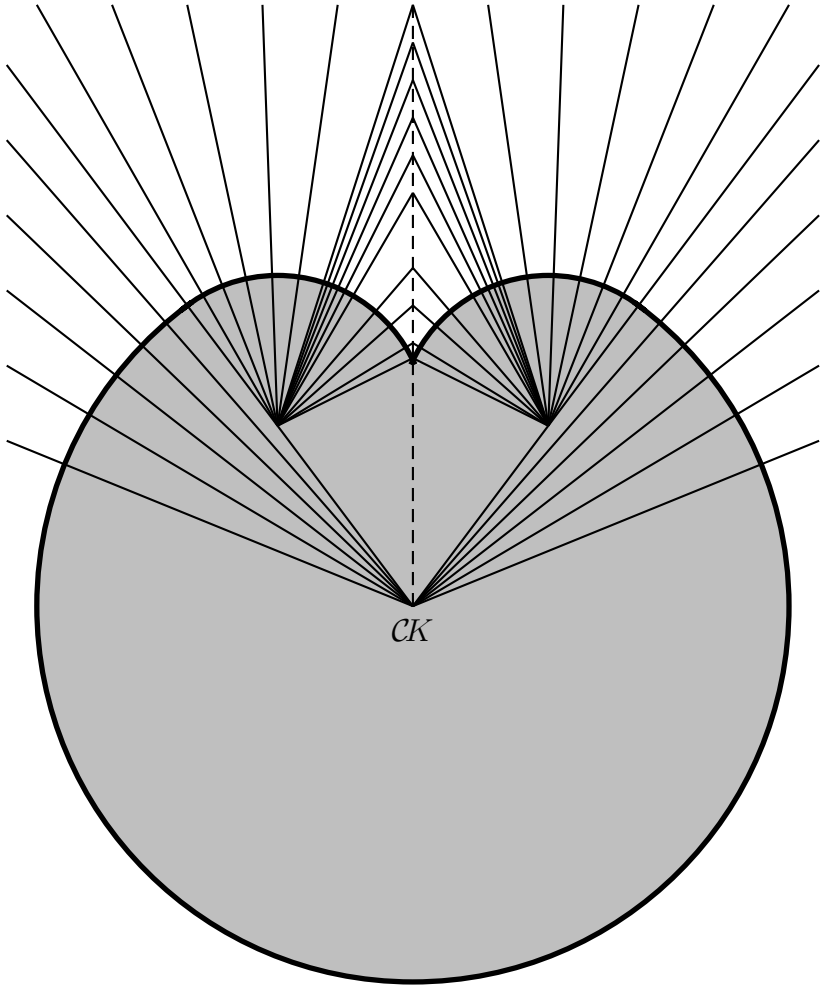

\pspicture(-5.5,-5)(5.5,8)
\pscustom[fillstyle=solid,fillcolor=lightgray,linewidth=2pt]{
\psarc[liftpen=2](-1.8,2.4){2}{24}{130}
\psarc[liftpen=1](0,0){5}{126}{54}
\psarc[liftpen=2](1.8,2.4){2}{50}{156}}
\psline[linestyle=dashed](0,0)(0,8)
\psline(-1.8,2.4)(0,8)
\psline(1.8,2.4)(0,8)
\psline(-1.8,2.4)(0,7.5)
\psline(1.8,2.4)(0,7.5)
\psline(-1.8,2.4)(0,7)
\psline(1.8,2.4)(0,7)
\psline(-1.8,2.4)(0,6.5)
\psline(1.8,2.4)(0,6.5)
\psline(-1.8,2.4)(0,6)
\psline(1.8,2.4)(0,6)
\psline(1.8,2.4)(0,5.5)
\psline(-1.8,2.4)(0,5.5)
\psline(1.8,2.4)(0,4.5)
\psline(-1.8,2.4)(0,4.5)
\psline(1.8,2.4)(0,4)
\psline(-1.8,2.4)(0,4)
\psline(1.8,2.4)(0,3.5)
\psline(-1.8,2.4)(0,3.5)
\psline(1.8,2.4)(0,3.3)
\psline(-1.8,2.4)(0,3.3)
\psline(1.8,2.4)(1,8)
\psline(-1.8,2.4)(-1,8)
\psline(1.8,2.4)(2,8)
\psline(-1.8,2.4)(-2,8)
\psline(1.8,2.4)(3,8)
\psline(-1.8,2.4)(-3,8)
\psline(1.8,2.4)(4,8)
\psline(-1.8,2.4)(-4,8)
\psline(1.8,2.4)(5,8)
\psline(-1.8,2.4)(-5,8)
\psline(0,0)(5.4,7.2)
\psline(0,0)(-5.4,7.2)
\psline(0,0)(5.4,6.2)
\psline(0,0)(-5.4,6.2)
\psline(0,0)(5.4,5.2)
\psline(0,0)(-5.4,5.2)
\psline(0,0)(5.4,4.2)
\psline(0,0)(-5.4,4.2)
\psline(0,0)(5.4,3.2)
\psline(0,0)(-5.4,3.2)
\psline(0,0)(5.4,2.2)
\psline(0,0)(-5.4,2.2)
\uput[270](0,0){${\cal C}\!K$}
\endpspicture
\caption{The portion of the solid segments that lies \emph{outside} of the
shaded area is the projection of the generators of ${\mathcal{H}}%
^+(K(a,r_1)) $ on the plane {$t=0$}.}
\label{projection}
\end{figure}

Consider now a set $K(a,r_1)$,
defined by the addition of a ripple to $%
S(0,1) $ at $y_0=\mathrm{e}^{i\frac{\pi}2}$, of radius $r_1$ and diameter
$a$, with $%
a<r_1<1/2$. (By that we mean that $K(a,r_1)$ is that set which has as boundary
$S(0,1)$ with a ripple added, and ${\mathbb R}^3\setminus B(0,1)\subset
K(a,r_1)$). From what has been said so far it follows that those generators
of ${\mathcal{H}}^+(K(a,r_1))$ which emanate from points lying on $S(0,1)$
project on the radial half lines $\{r\mathrm{e}^{i\varphi},r\in [1,\infty)\}$%
. Those generators of ${\mathcal{H}}^+(K(a,r_1))$ which emanate from points $%
\novec x$ lying on $S(\novec x_1,r_1)
\setminus\{r\mathrm{e}^{i\frac{\pi}2},r\in{\mathbb R}\}$ project to line
segments contained in the lines $\{r(\novec x-\novec x_1),r\in{\mathbb R}%
\} $. Note that each such segment, except for the point $y_1$ where $%
S(\novec x_1,r_1)$ meets $S(0,1)$, intersects the axis
$\{r\mathrm{e}^{i\frac{\pi}2},r\in{ \mathbb R}\}$ precisely at one point. By
invariance of $K(a,r_1)$ under reflections across that axis it follows
that each such generator must have an end point precisely on the plane
$\{t\in{\mathbb R}, \novec x=r\mathrm{e}^{i\frac{\pi}2}, r\in{\mathbb R}\}$. The
projection of some of the generators of ${ \mathcal{H}}^+(K(a,r_1))$
can be found in Figure~\ref{projection}.

We wish to point out the following:

\begin{enumerate}
\item  The generators can always be parameterized so that they are given by
an equation of the form $\gamma =\{x^\mu
(s),\dot{x}^0=1,|\dot{x}|=1\}$, with $%
x^\mu (0)\in \partial K(a,r_1)$. In everything that follows all generators
will always be parameterized in that manner.

\item  Let $C={\mathcal{H}}^{+}(K(a,r_1))\cap \{(t,\novec{x}),t\in {\mathbb R},%
  \novec{x}=r\mathrm{e}^{i\frac{\pi}2 },r\in {\mathbb R}\}$. Every point $p\in
  C$ is the future end point of two generators of
  ${\mathcal{H}}^{+}(K(a,r_1))$, while every point $p\in
  {\mathcal{H}}^{+}(K(a,r_1))\setminus C$ is an interior point of
  precisely one generator of ${\mathcal{H}}^{+}(K(a,r_1))$. $C$ will
  be called the crease set of ${\mathcal{H}}^{+}(K(a,r_1))$.
\end{enumerate}

\begin{figure}[thbp]
\includegraphics[width=\textwidth]
{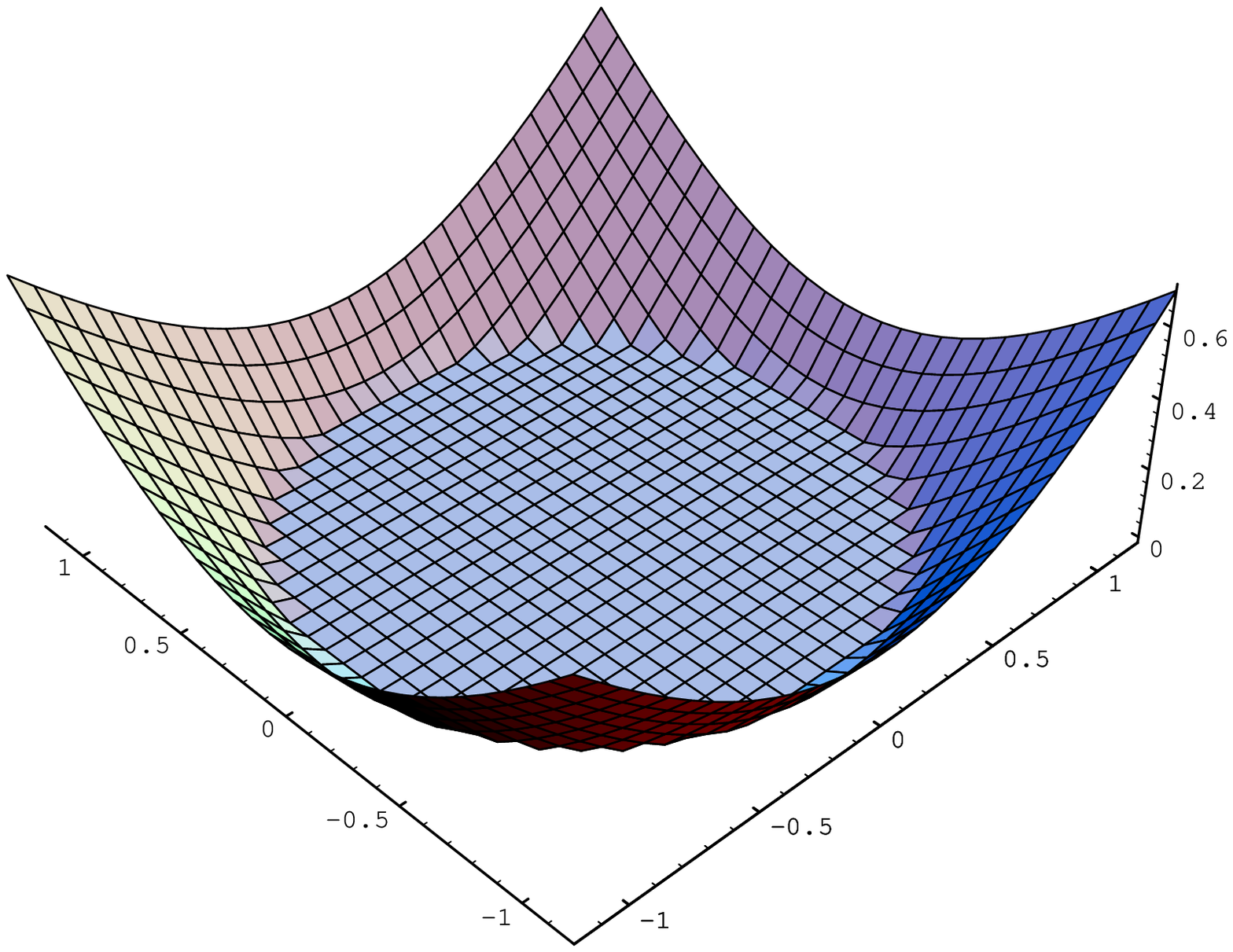}
\caption{The future Cauchy horizon $\mathcal{H}^+(K_0) $ of $K_0 $.}
\label{zerothCauchyhorizon}
\end{figure}

We wish now to set up an iterative scheme for constructing a sequence
$(K_i, \varepsilon_i,\delta_i)$, where $\varepsilon_i$ and $\delta_i$
are positive numbers and $K_i$ is an increasing sequence of sets. The
reader is referred to Section \ref{S4} for the definition of
semi--tangents and of the distribution of semi--tangents. For any set
$A\in{\cal X}$, with ${\cal X}$ as in \eq{Xidef}, we define \emph{the crease
    set} of $\mathcal{\ H}^+(A) $ as the set of points of
$\mathcal{\ H}^+(A) $ at which there is more than one semi-tangent. We set
  $K_0={\mathbb R} ^3\setminus B(0,1)$, $\varepsilon_0=\infty$,
  $\delta_{0}=\infty$.
  The future Cauchy horizon $\mathcal{\ H}^+(K_0) $ of $K_0 $ is shown
  in Figure \ref{zerothCauchyhorizon}.  The number $\varepsilon_i$
  will be a measure of ``how close to each other are distinct
  semi--tangents on the crease set of ${\mathcal{H}}^+(K_i)$''.  The
  number $\delta_i$ will be a measure
of ``how far apart are the distributions of semi--tangents ${\mathcal{N}}%
{}^+K_{i-1} $ and ${\mathcal{N}}{}^+K_i$''. Clearly the crease set $C_0$ of $%
K_0$ is empty.

\begin{figure}[thbp]
\centering
\includegraphics[width=0.6\textwidth]
{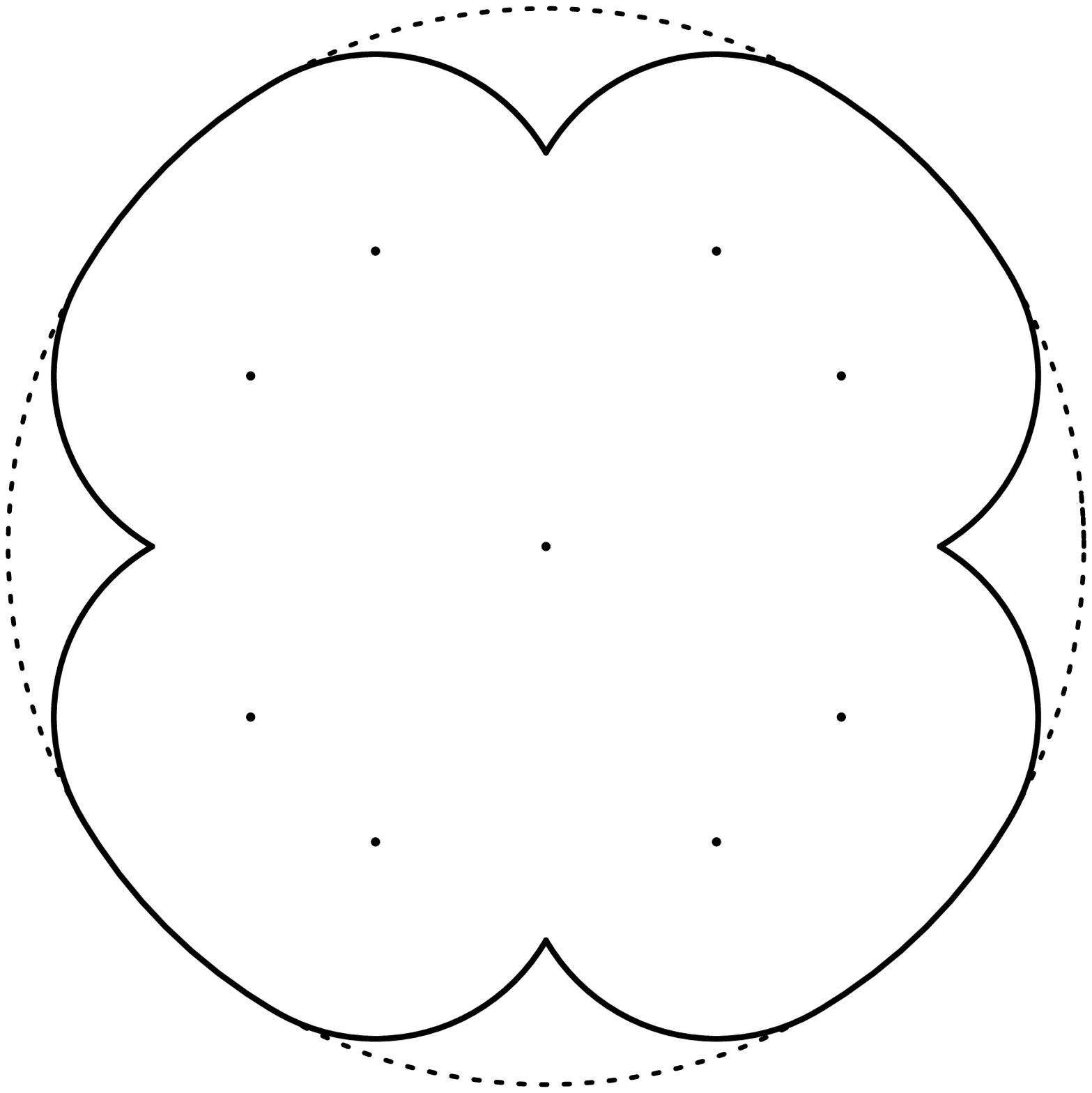}
\caption{The dotted line is the boundary of $K_0$, the solid line that of $%
  K_1$. For graphical clarity the parameters describing the ripples
  have {\em not} been chosen so that the modulus of Lipschitz
  continuity of the function representing $K_1$ be smaller than one half,
  as required in the construction described below.}
\label{firststep}
\end{figure}
\begin{figure}[thbp]
\includegraphics[width=\textwidth]
{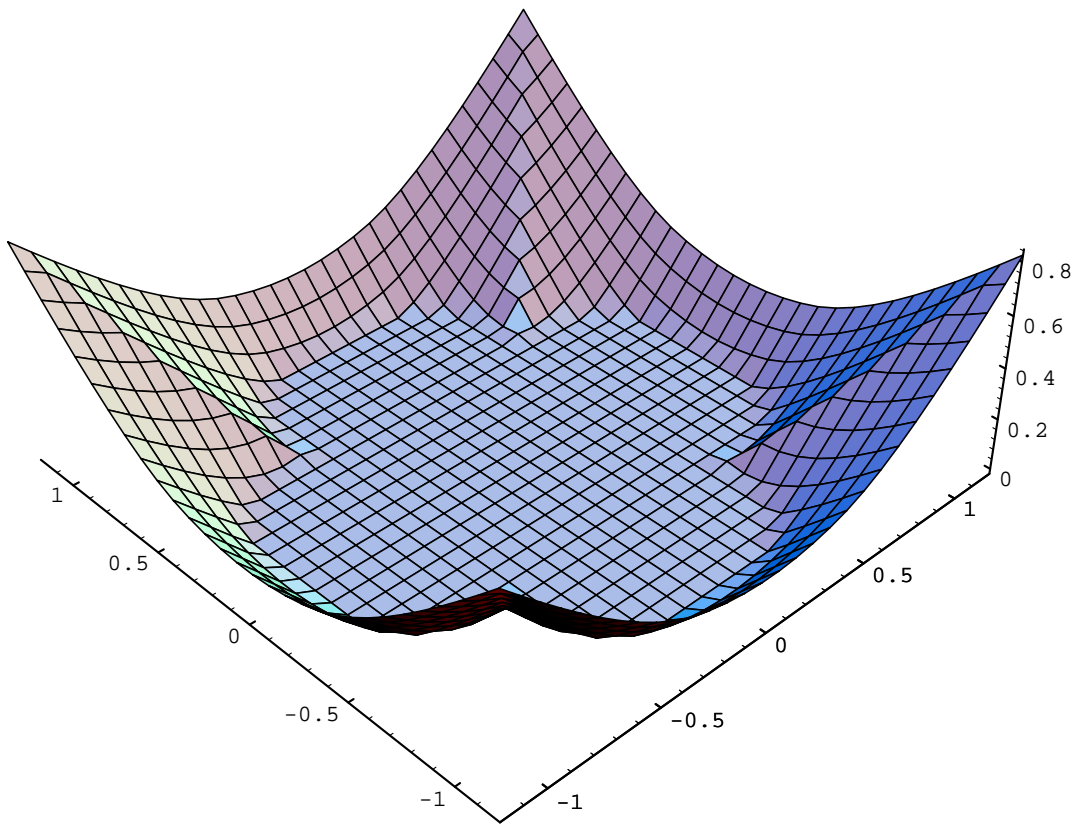}
\caption{The future Cauchy horizon $\mathcal{H}^+(K_1) $ of $K_1
  $. ($K_1$ here has been rotated by $45$ degrees, as compared to
  Figure \ref{firststep}.)}
\label{secondCauchyhorizon}
\end{figure}

Let $\{y_n=2\mathrm{e}^{\frac{i\pi n}2},n=1,\ldots ,4\}$, and let $x_n$ be
the starting points of those generators of ${\mathcal{H}}^{+}(K_0)$ which
pass through $y_n$, thus $x_n=\mathrm{e}^{\frac{i\pi n}2}$.
We define $K_1$
by adding a ripple of radius $r_1$ and diameter $a_1$ at each $x_n$. $r_1$
and $a_1$ are arbitrary except for the requirement that the ripples do not
intersect, and that $\partial K_1$ be the radial graph of a function $f_1$, $%
\partial K_1=\{r=f_1(\theta ),\theta \in [0,2\pi ]\}$, such that the modulus
of Lipschitz continuity of $f_1$ is smaller than $1/2$. A possible set $K_1$
has been shown in Figure \ref{firststep}. By the localization principle,
Lemma \ref{L3.7}, the horizon ${\mathcal{H}}^{+}(K_1)$ will coincide with
that of ${\mathcal{H}}^{+}(K_0)$ away from those circular arcs at which ripples
have been added. Again by the localization principle in a small neighborhood
of each ripple the horizon will look like the horizon of the set $K(a,r_1)$ as
previously considered -- ${\mathbb R}^3\setminus B(0,1)$ with a single
ripple added. Thus ${\mathcal{H}}^{+}(K_1)$ will look as shown in
Figure~\ref{secondCauchyhorizon}.
It follows that the crease set $C_1$ of
${\mathcal{H}}^{+}(K_1)$ projects down on ${\mathbb R}^2$ under $\Pi $
to half--infinite intervals included in the
four half-lines $\{r\mathrm{e}^{i%
  \frac{\pi n}2},r\in [1,\infty )\}$, $n=1,\ldots ,4$. It should be
noted that every point in the annulus $\overline{B(0,2)\setminus
B(0,1)}$ is a distance not further than $\frac \pi 2$
from $\Pi C_1$.
Letting $\rho $ be the distance on $T{\mathbb R}^{2,1}$ as defined
after the example \ref{e5.2} in Section \ref{S3}, we set
\[
\varepsilon _1=\inf \{\rho (X,Y)|\;X,Y\in {\mathcal{N}}{}_pK_1,X\neq Y,p\in
C_1\},
\]
\[
\delta _1=\varepsilon _1/4.
\]
This finishes our first inductive step.

\begin{figure}[htbp]
\begin{center}
\leavevmode
\includegraphics[width=0.5\textwidth]
{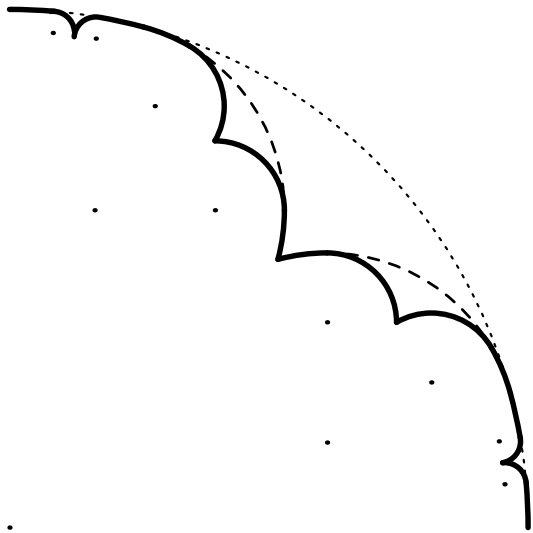}
\end{center}
\caption{The dotted line represents the intersection of the boundary of $K_0$
with
the first quadrant, the dashed one that
of $K_1$, and the solid one that
of $K_2$. The dots are the centers of the corresponding circles.
$K_0$ and $K_1$ have been rotated by $45^o$, as compared to
Fig.~\ref{firststep}.
For graphical clarity the radii of the
``second-generation-ripples'' have not been chosen to be equal, as
done in the construction below.}
\label{secondstep}
\end{figure}

The remainder of the construction consists in adding ripples in a
controlled way. The second step is shown in Figure \ref{secondstep},
and that figure makes it also clear how the induction proceeds; the
formal description goes as follows: Suppose that $(K_j, \varepsilon_j,
\delta_j) $ have been constructed for all $j\leq i$, with the property that $%
\Pi C_j\subset \Pi C_{j+1}$, and
\[
\delta_{j+1}\leq \frac{1}{4}\min(\varepsilon_j,\delta_j),
\]
and with the property that for $j\ge 1$ every point $p\in
\overline{B(0,2)\setminus B(0,1) }$ is a distance not further than
$2^{-j}\pi$ from $\Pi C_j$. Let us further assume that for
$j=1,\ldots,i$ the $K_j$'s have the property that $\partial K_j$ is a
radial graph $\partial K_j=\{ r=f_j(\theta),\theta\in [0,2\pi]\}$ of a
Lipschitz continuous function, the modulus of continuity of which is
smaller than $\frac12+\frac{1}{2^2}+\ldots+\frac{1}{2^j}$. We construct $%
K_{i+1}$ as follows: we can find points
$z_k\in
C_i \cup ({\mathcal{H}}^+(K_i) \cap\Pi^{-1}(\partial B(0,2))$
$k=1,\ldots,I$, such that
every point $p\in {\mathcal{H}}^+(K_i) \cap
\Pi^{-1}(\overline{B(0,2)})$ is a distance not larger than
$2^{-i-1}\pi$ from a generator of ${\mathcal{H}}^+(K_i)$ passing
through one of the $z_k$'s, or having its end point there.  Let
$\{y_n\}_{n=1}^I$ denote the set of the starting points of those
generators. We may choose the $z_k$'s so that none of the $y_n$'s lies
on
the end point of a circular arc on $\partial K_i$. We can also require that $%
\# {\mathcal{N}}{}_{z_k} K_i\le 2$.


Define $K_{i+1}(a)$ by adding ripples of radius $r_{i+1}$ and diameter $a$, $%
a\leq a_{i+1}\leq r_{i+1}$, to $\partial K_i$ at  all the $y_n$'s. The
radius $r_{i+1}$
and threshold diameter $a_{i+1}$ are chosen small enough so that $\partial
K_{i+1}$ is a radial graph of a function $f_{i+1}$, the modulus of Lipschitz
continuity of which is smaller than or equal to $\frac12+ \frac{1}{2^2}%
+\ldots+\frac{1}{2^{i+1}}$. Moreover $a_{i+1}$ is chosen small enough so
that the new ripples do not intersect each other. Consider now a point $%
z_k\in C_i$, such that $z_k\in C_j$ and $z_k\notin C_{j-1}$ for some $j$.
There exist precisely two points, say $x_1$ and $x_2$, on $\partial K_i$,
which are initial points of generators of ${\mathcal{H}}^+(K_i)$, say $%
\gamma_1$ and $\gamma_2$, with end point at $z_k$.
Moreover $x_1$ and $x_2$
are symmetric images of each other under the reflection across the line
which contains the projection of that generator $r_k$ of ${\mathcal{H}}%
^+(K_{j-1})$ which passes through $z_k$. By the localization principle,
Lemma \ref{L3.7}, the addition of a sufficiently small ripple at $x_1$ and $%
x_2$ will only affect ${\mathcal{H}}^+(K_i)$ in a small neighborhood of $%
\gamma_1$ and $\gamma_2$. Because the ripples are added symmetrically
with respect to $r_k$, the generators emerging from points on the
ripples can only meet on a set, the projection of which contains the
projection of $r_k$. It follows that for $a_{i+1}$ small enough the
projection on ${\mathbb R}^2$ of the crease set $C_{i+1}(a)$ of
${\mathcal{H}}^+(K_{i+1}(a))$, $a \le a_{i+1}$, will contain the projection
of the
crease set $C_i$.

Note that the above discussion also shows that there exists a compact
neighborhood $D_i$ of $\{z_k\}$, with $D_i\cap \partial
K_i=\emptyset$, such that the crease set of ${\mathcal{H}}^+( K_i) $
is contained in that of ${\mathcal{H}}^+(K_{i+1}(a))$, except for points
in $D_i$, for all $0\leq a<a_{i+1}$.
Moreover, by choosing $a_{i+1}$ sufficiently small, $D_i$ may be expressed
as a disjoint union,
$D_i =
\cup_k D_{i,k}$, where $D_{i,k}$ is an arbitrarily small compact
neighborhood of $z_k$.
Then,
continuity properties of geodesics, as applied to the null geodesics
orthogonal to
$\partial  K_{i+1}$,
and the structure of the crease sets imply the following: For any $\epsilon
> 0$, $a_{i+1}$
(and, in turn, the $D_{i,k}$'s) may
be chosen sufficiently small so that for any
$Z \in \mathcal N^+\!K_{i+1}(a) \cap \hat \Pi^{-1}(\Pi D_{i,k})$ and for any
$q \in \mathcal H^+(K_{i+1}(a)) \cap \Pi^{-1}(\Pi (C_i \cap D_{i,k}))$
there exists
$Y \in \mathcal N^+_q\!K_{i+1}(a)$ such that $\rho(Z,Y) < \epsilon$.

Define
\begin{equation}  \label{Eq.1}
\delta_{i+1}{(a)} =\sup\{\inf\{ \rho(X,Y)|\; Y\in {\mathcal{N}}%
{}_{\Pi^{-1}p}^+ K_{i+1}(a) \} |\; p\in \Pi C_i,X\in{\mathcal{N}}%
{}_{\Pi^{-1}p}^+ K_i\}.
\end{equation}
Here $\Pi^{-1}p$ denotes that point which is the lift of $p$ to the
appropriate Cauchy horizon.
By our previous remark, in (\ref{Eq.1}) we
can replace the condition $p\in \Pi C_i$ by: $p\in \Pi(C_i \cap D_i)$,
where $D_i$
is as above.
Hence, from our previous remark and part (b) of
Theorem
\ref{T5.3} it follows that $ \delta_{i+1}(a) \rightarrow 0$ as
$a\rightarrow0$. (This should
anyway be clear by inspecting what happens in Figures \ref{projection}
or \ref{zerothCauchyhorizon} when the diameter of the ripples tends to
zero). We can thus choose $a_{i+1}$ small enough so that
\[
\delta_{i+1}(a_{i+1})\le\frac{1}{4}\min({\delta_i,\varepsilon_i}).
\]
Set $K_{i+1}=K_{i+1}(a_{i+1})$, and define
\[
{\varepsilon}_{i+1}=\inf\{\inf\{ \rho(X,Y) \,|\; X,Y\in{\mathcal{N}}%
{}_{\Pi^{-1}p}^+ K_{i+1}, X \ne Y\}\,|\; p\in C_{i+1}\} > 0,
\]
where, $C_{i+1}$ is the crease set of ${\mathcal{H}}^+(K_{i+1})$.
This completes our inductive step.

Consider the family of functions $f_i(\theta)$ which represent
$\partial K_i$ as radial graphs. For any $\theta$ we have $f_{i+1}
(\theta)\leq f_i(\theta)$ , so that $f(\theta)=\displaystyle
\lim_{i\rightarrow\infty} f_i(\theta)$ exists. Set
\[
K=\{r\mathrm{e}^{i\theta} |\; r\geq f(\theta)\}.
\]
As the modulus of Lipschitz continuity of all the $f_i$'s is less than $%
\frac12+\frac14+\ldots+ \frac{1}{2^i}<1$, $f$ is a Lipschitz continuous
function and $\partial K$ is thus a Lipschitz continuous topological
manifold.

Let $C$ be the crease set of ${\mathcal{H}}^+(K)$, we wish to show that $\Pi
C$ is dense in the annulus $\overline{B(0,2)\setminus B(0,1)}$. This will
follow if we show that $\Pi C$ contains $\Pi C_i$ for each $i$.

Consider then a point $p\in C_i$, there exist at least two vectors $X$ and $%
Y $ in $T M$ which are semi-tangent to ${\mathcal{H}}^+(K_i)$, with $%
\rho(X,Y)\geq \varepsilon_i$, let $q=\Pi p$. By construction there
exist $X_1, Y_1\in {\mathcal{N}}{}_{p_1}^+ K_{i+1}$, $\Pi p_1=\Pi p$,
such that $ \rho(X,X_1)\leq\delta_{i+1}\leq \frac{\varepsilon_i}{4}$,
$\rho(Y,Y_1)\leq \frac{\varepsilon_i}{4}$. Similarly there exist
$X_2,Y_2\in{\mathcal{N}}
{}_{p_2}^+ K_{i+2}$ such that $\rho(X_1,X_2)\leq\frac{\delta_{i+1}}{4}%
\leq\frac{\varepsilon_i}{16}$, $\rho(Y_1,Y_2)\leq\frac{\varepsilon_i}{16} $.
This gives $\rho(X,X_2)\leq \frac{\varepsilon_i}{4} + \frac{\varepsilon_i}{16%
}<\frac{\varepsilon_i}{3}$, $\rho(Y,Y_2)<\frac{\varepsilon_i}{3}$.
Continuing this way one obtains a sequence of vectors $X_i$, $Y_i$ such that
$\rho(X,X_i)<\frac{\varepsilon_i}{3}$, $\rho(Y,Y_i)<\frac{\varepsilon_i}{3}$%
. By compactness a converging sequence can be chosen, converging to vectors $%
X_{\infty}$, $Y_{\infty}$. By the triangle inequality we have $%
\rho(X_\infty,Y_\infty)\geq\frac{\varepsilon}{3}$, so that $X_\infty\neq
Y_\infty $. By the cluster-limit Lemma \ref{L5.1}, $X_\infty,Y_\infty
\in {\mathcal{N}}{}_{\Pi^{-1}q}^+ K$. By Proposition \ref{P4.4} ${\mathcal{H}%
}^+(K)$ cannot be smooth at $\Pi^{-1}q$, and the result follows. \hfill\ $%
\Box $

\section{Higher dimensional examples}

\label{ex}

Let us now address the question of existence of ``nowhere'' differentiable
horizons in more than $2+1$ space--time dimensions. First, consider the
space--times $M^{p,q}={\mathbb R}^{2,1}\times{\mathbb R}^p\times S_1^q$,
where $S_1$ denotes a circle, $p$, $q$ are non-negative integers. Here ${%
\mathbb R}^p$ and $S_1^q$ are equipped with the obvious metric, and $M^{p,q}$
is equipped with the product metric. We claim that
\begin{equation}  \label{7.1}
{\mathcal{H}}^+(K\times{\mathbb R}^p\times S_1^q;M^{p,q})= {\mathcal{H}}^+(K;%
{\mathbb R}^{2,1})\times{\mathbb R}^p\times S_1^q\ .
\end{equation}
Here $K$ is the set constructed in Theorem \ref{T1}, ${\mathcal{H}}%
^+(\Omega;M)$ denotes the Cauchy horizon of an acausal set $\Omega$ in a
space--time $M$. It follows that ${\mathcal{H}}^+(K\times{\mathbb R}^p\times
S^q;M^{p,q})$ provides us with a ``nowhere'' differentiable Cauchy horizon,
via the constructions discussed in Section 1. To prove (\ref{7.1}) it is
sufficient to note that ${\mathbb R}^p\times S_1^q$ acts transitively on
itself by isometries, which consist of translations in the ${\mathbb R%
}^p$ factor and rotations of each $S_1$ factor. Since ${\mathcal{H}}%
^+(\Psi(\Omega);M)=\Psi({\mathcal{H}}^+(\Omega;M))$ for any isometry
$\Psi$, it follows that ${\mathcal{H}}^+(K\times{\mathbb R}^p\times
S^q;M^{p,q})$ is of the form $\mathcal{U}\times{\mathbb R}^p\times
S^q$ for some set $\mathcal{U}
$. Clearly $\mathcal{U}={\mathcal{H}}^+(K;{\mathbb R}^{2,1})$ and (\ref{7.1}%
) follows.

Note that if $p>0$ then $K\times{\mathbb R}^p\times S^q$ does not have a
compact boundary, while if $q>0$ then $M^{p,q}$ is not asymptotically flat
in the usual sense. It is therefore of interest to try to mimic our
construction in higher dimension, to obtain a set $K\subset {\mathbb R}^n$,
with compact boundary $\partial K$, having the properties listed in Theorem
\ref{T1}, when ${\mathbb R}^n$ is considered as the $t=0$ hyper-surface in $%
n+1$-dimensional Minkowski space--time ${\mathbb R}^{n,1}$. Now the results
proved in Sections \ref{S3} and \ref{S4} are dimension-independent, let us
then isolate the essential elements of our construction.

The key element was the procedure of adding a ripple to the boundary
of a set, which then produced a crease on the event horizon. Thus we
need a building block which will have an effect similar to the one
that our two-dimensional ripple had. Such a higher dimensional ripple
can be easily constructed as follows: let us start with $3+1$
dimensions. Let $K(a,r)$ be the complement of a disc with a ripple
added, as described at the beginning of Section 2 and shown in
Figure~\ref{projection}, lying in the plane $z=0$ in ${\mathbb R}^3$.
Let $\hat{K} (a,r)\subset {\mathbb R}^3$ be the set obtained by
rotating $K(a,r)$ around the axis $z=x=0$. ($K(a,r)$ is clearly
diffeomorphic to an apple without its stalk.) $\hat{K}(a,r)$ is
invariant under rotations around that axis, so must therefore be its
Cauchy horizon in ${\mathbb R}^{3,1}$.
It is then easily seen that ${\mathcal{H}}^+\equiv {\mathcal{H}}^+(\hat{K}%
(a,r);{\mathbb R}^{3,1})$ is obtained by rotating ${\mathcal{H}}^+(K(a,r);{%
\mathbb R}^{2,1})$ around the $z=x=0$ axis. This shows that the crease set
of ${\mathcal{H}}^+$ is non-trivial and projects on ${\mathbb R}^3$ under
the canonical projection $\Pi:{\mathbb R}^{3,1}\rightarrow{\mathbb R}^3$ to
a subset of the $z=x=0$ axis. Clearly one can now start an iterative scheme
of adding new ripples, and obtain a desired set $K$ by passing to a limit.

Let us mention that a useful bookkeeping procedure in
the
two-dimensional case was to ensure that the projections $\Pi C_i$ of
the crease sets $C_i$ formed an increasing sequence of sets.
Presumably a way of ensuring that could also be conceived in the
three-dimensional case. Rather than doing that let us note that this
aspect of the construction is not necessary, and can be bypassed as
follows:
At the $i$-th step of the construction add the new ``end
points'' $\{z_j\}_{j=1}^{N_i}$ of the generators $\gamma_j$,
 $j=1,\ldots,N_i$, so that every point of
$\overline{ B(0,2)\setminus B(0,1)}\subset {\mathbb R}^3$ be a
distance not larger than $ 2^{-i-2}$ from the projection of some
generator with future end point in the set $ \{z_j\}_{j=1}^{N_i}$,
 or passing through a point in $ \{z_j\}_{j=1}^{N_i}$. Choose
then the size of the ripples added at the $z_j$'s, $j=1,\ldots,N_i$ so
small that $\Pi C_{i+1}$ lies a
distance not larger than $2^{-i-2}$ from one of the distinguished
generators $\gamma_i$
or from $\Pi C_i$.
The other arguments go through with some rather
obvious modifications.

Let us finally describe one of the many possible processes of ``adding a
ripple'' in $n$-dimension. Let $f_{a,r}$ be the function which describes the
``rippled circle'' $\partial K(a,r)$ as a radial graph, $\partial
K(a,r)=\{f_{a,r}(\varphi )\mathrm{e}^{i\varphi}, \varphi\in[0,2\pi]\}$; $%
K(a,r)$ is as described at the beginning of Section 2. Then a ``rippled $n-1$%
-dimensional sphere'' $\partial \tilde{K}(a,r)$ is given by the equation
\begin{eqnarray*}
\partial \tilde K(a,r)=& \{  x_n=f_{a,r}(\theta)\cos\theta,
\qquad \qquad \qquad
\qquad \qquad \qquad \qquad
\\
 &\qquad
x_1^2+x_2^2+\ldots+x_{n-1}^2=f_{a,r}(\theta)\sin\theta, \qquad
\theta\in[-\frac{\pi%
}{2},  \frac{\pi}{2}]\}.
\end{eqnarray*}

The reader should be able to check, by isometry considerations, that for all
$t\geq0$ the cross-sections of ${\mathcal{H}}^+\equiv {\mathcal{H}}^+(\tilde
K(a,r);{\mathbb R}^{n,1})$ with the hyper-planes $t=$ const are sets the
boundaries of which are again rippled $n-1$ spheres. The crease set of ${%
\mathcal{H}}^+$ projects down to a subset of the axis $x^1=x^2=%
\ldots=x^{n-1}=0$.

\noindent
\textbf{Acknowledgments} P.T.C.  acknowledges useful discussions
with R. Geroch, K. Newman, P. Tod and R. Wald. He also wishes to thank the
Department of Mathematics of University of Miami for hospitality during part
of work on this paper. We are grateful to V.E. Dubau and C. Georgelin for
producing the {\sc Mathematica} figures.

\bibliographystyle{/usr/local/lib/texmf/bibtex/bst/amsplain}
\bibliography{$HOME/prace/references/hip_bib,%
$HOME/prace/references/reffile,%
$HOME/prace/references/newbiblio}

\end{document}